\documentclass[letterpaper,12pt]{article}
\usepackage{jheppub}
\usepackage{subcaption}
\usepackage{bm}
\allowdisplaybreaks

\def\sgn{\text{sgn}}
\def\e{\epsilon}
\def\r{\rho}
\def\p{\pi}
\def\f{\phi}

\def\b{\beta}
\def\h{\eta}
\def\t{\tau}
\def\s{\sigma}
\def\q{\theta}
\def\pa{\partial}
\def\bal#1\eal{\begin{align}#1\end{align}}

\title{Soft modes in $\mathcal{N} = 2$ SYK model}
\author{Cheng Peng${}^{a,b}$, and Stefan Stanojevic${}^b$}
\affiliation{${}^a$ \it Center for Quantum Mathematics and Physics (QMAP), Department of Physics\\
\qquad 	University of California, Davis, CA 95616 USA\\
\vspace{0.2cm}  
${}^b$ Department of Physics, Brown University, Providence RI 02912, USA}

\emailAdd{cpeng@ucdavis.edu}
\emailAdd{stefan\_stanojevic@brown.edu}

\abstract{We study various properties of the soft modes in the $\mathcal{N}=2$ supersymmetric SYK model. }

\begin{document}
\maketitle

\section{Introduction}
The Sachdev-Ye-Kitaev model~\cite{Sachdev:1992fk,Parcollet:1997ysb,KitaevTalk1,KitaevTalk2,Polchinski:2016xgd,Maldacena:2016hyu,Kitaev:2017awl,Jevicki:2016bwu,Maldacena:2016upp} motivates the recent advances in holographic understanding of (quantum) gravity~\cite{Kourkoulou:2017zaj,Saad:2018bqo,Penington:2019npb,Almheiri:2019hni,Saad:2019lba,Almheiri:2019qdq,Penington:2019kki}. The SYK model admits supersymmetric generalizations~\cite{Fu:2016vas}. In this paper we focus on the $\mathcal{N}=2$ supersymmetric SYK model. Some details of the correlation functions are computed in~\cite{Peng:2017spg}. The partition function of this model is discussed in detail in~\cite{Stanford:2017thb, Mertens:2017mtv}. Similar to its purely fermionic counterpart~\cite{Cotler:2016fpe,Berkooz:2018qkz}, the supersymmetry model can also be studied in the doubly scaled large-$q$ limit~\cite{Berkooz:2020xne}. Some properties of the spectrum of the model are discussed in~\cite{Kanazawa:2017dpd}. A bulk interpretation of the supersymmetric Schwarzian model is discussed in~\cite{Forste:2017apw}.  Supersymmetry turns out to be crucial in the construction of higher dimensional covariant analogue of the disordered SYK model~\cite{Murugan:2017eto,Bulycheva2018,Peng:2018zap,Ahn:2018sgn,cepm}.  

In this work we continue the study of the correlation functions in the $\mathcal{N}=2$ SYK model~\cite{Peng:2017spg}, focusing on the low energy modes. Detailed discussion about these soft modes can be found in e.g.~\cite{Maldacena:2016hyu,Jevicki:2016ito,Kitaev2017,Bagrets2017,Mertens:2017mtv,Das1802571} After a brief review of the $\mathcal{N}=2$ SYK model,  we start with an analysis of the large-$q$ limit. In particular, we work out the large-$q$ propagators in section~\ref{sec:largeqpropagators}, the corrections to the eigenvalues of the ``nondiagonal" kernels are found in section~\ref{sec:nond}, and the corrections to the ``diagonal" kernels in~\ref{sec:appdiagonal}.  We evaluate Lyapunov exponents in section~\ref{sec:chaos}. We further discuss the effective action of the soft modes corresponding to the spontanuous and explicit breaking of the super-reparameterization in section~\ref{sec:schwarzian}. We discuss the contribution of the exact ground states to the correlation function in section~\ref{gd} and show that their contribution is negligible at slightly higher temperature so the full correlator at finite temperature could be obtained from the conformal part of the zero temperature correlation function by a reparameterization. Finally we consider the correlators of the Schwarzian operators in section~\ref{SC}.

\section{Review of the $\mathcal{N}=2$ supersymmetric SYK model}
\label{sec:kernels}

The Lagrangian of the $\mathcal{N} = 2$ SYK model reads~\cite{Fu:2016vas,Peng:2017spg}
\bal\label{action}
L = \bar{\psi}_i\pa_\t \psi_i - \bar{b}_i b_i+ i^{(q-1)/2} C_{i \, j_1 \, \hdots \, j_{q-1}} \bar{b}_i \psi_{j_1} \hdots \psi_{j_{q-1}} + i^{(q-1)/2} \bar{C}_{i j_1 ... j_{q-1}} {b}_i \bar{\psi}_{j_1} ... \bar{\psi}_{j_{q-1}}\,,
\eal
where the Gaussian distribution of random coupling satisfies
\bal
\langle C_{i_1 \hdots i_q} \bar{C}_{i_1 \hdots i_q} \rangle = \frac{(q-1)! J}{N^{q-1}}\ .\label{GaussJ}
\eal

The model is proposed in~\cite{Fu:2016vas}. Four point correlation functions in the $\mathcal{N} = 2$ SYK model are computed explicitly in \cite{Peng:2017spg}. The connected piece of the four-point function, at the leading order of $\frac{1}{N}$, only receives contributions from the set of the ladder diagrams that can be iteratively generated by the action of a set of ladder kernels. In the $\mathcal{N} = 2$ SYK model, there are several kinds of relevant kernels.

\begin{figure}
\centering
\includegraphics[width=0.8\textwidth]{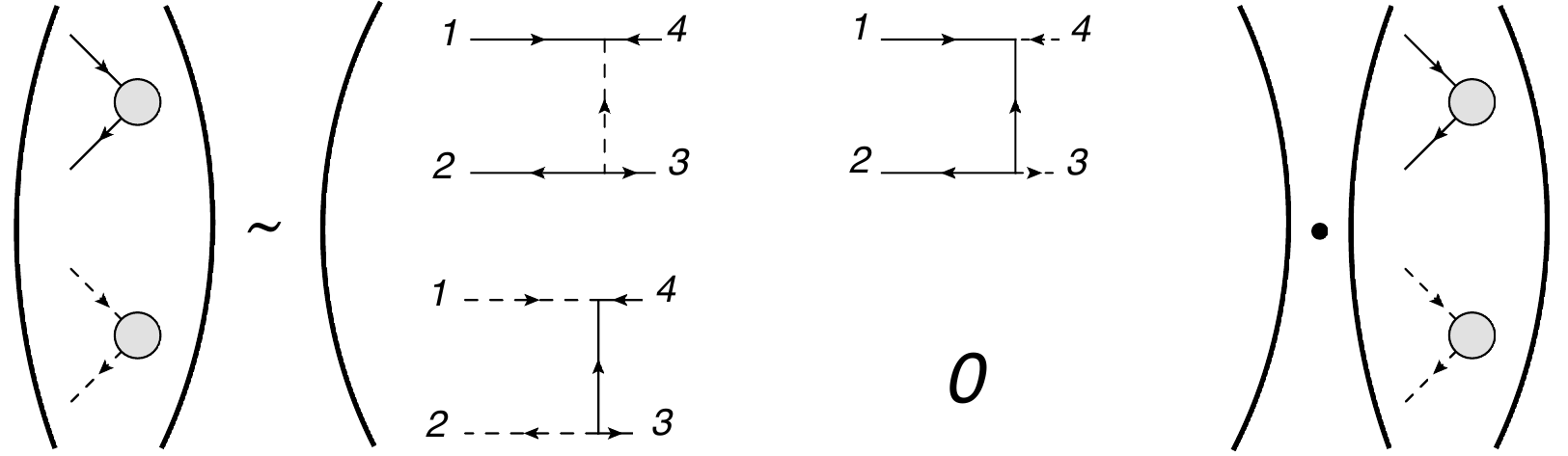}
\caption{ Action of the non-diagonal kernels $K^{11}$, $K^{12}$, $K^{21}$. In the plots the solid lines represent fermionic propagators and the dashed lines represent bosonic propagators.}\label{fig:kmat}
\end{figure}

\begin{enumerate}
\item The correlation function  $\langle\psi_i(\t_1)b_i(\t_2)\bar\psi_j(\t_3)\bar{b}_j(\t_4)\rangle$ receives contributions from a set of ladder diagrams with a bosonic line and a fermionic line on the ladder rails. They can be constructed by repeated actions of the ''diagonal" kernel
\bal
\label{eq:diagonalk}
K^{d} = J (q - 1) G^{\psi}(\t_{1 4}) G^{b}(\t_{23}) \big( G^{\psi}(\t_{3 4}) \big)^{(q - 2)}\ .
\eal
In the conformal limit, the eigenvalues of this kernel are $k_c^{s,d}$ and $k_c^{a,d}$~\cite{Peng:2017spg}. 

\item The $\langle\psi_i(\t_1)\bar{\psi}_i(\t_2)\psi_j(\t_3)\bar{\psi}_j(\t_4)\rangle$, $\langle\psi_i(\t_1)\bar{\psi}_i(\t_2)b_j(\t_3)\bar{b}_j(\t_4)\rangle$, $\langle b_i(\t_1)\bar{b}_i(\t_2)\psi_j(\t_3)\bar{\psi}_j(\t_4)\rangle$ and $\langle b_i(\t_1)\bar{b}_i(\t_2)b_j(\t_3)\bar{b}_j(\t_4)\rangle$    correlation functions receive contributions from ladder diagrams with 2 bosonic lines or 2 fermionic lines on both rails. They can be constructed by repeated actions of the matrix of ``non-diagonal" kernels, as shown in Figure~\ref{fig:kmat}. The different components of this matrix are
\bal
\label{eq:nondiagonal}
K^{1 1} &= J \frac{(q-1)!}{(q-3)!} G^{\psi}(\t_{1 4}) G^{\bar{\psi}}(\t_{2 3}) G^{\bar{b}}(\t_{3 4}) \big( G^{\psi}(\t_{3 4}) \big)^{q-3}  \\
K^{1 2} &= J \frac{(q-1)!}{(q-2)!} G^{\psi}(\t_{1 4}) G^{\bar{\psi}}(\t_{2 3}) \big( G^{\psi}(\t_{3 4}) \big)^{q-2} \\
K^{2 1} &= J \frac{(q-1)!}{(q-2)!} G^{\bar{b}}(\t_{1 4}) G^{b}(\t_{2 3}) \big( G^{\psi}(\t_{3 4}) \big)^{q-2}
\eal
In the conformal limit, they have the eigenvalues $k_c^{s,1 1}(h)$, $k_c^{s,1 2}(h)$, $k_c^{s,2 1}(h)$ that correspond to symmetric eigenfunctions, and the eigenvalues $k_c^{a,1 1}(h)$, $k_c^{a,1 2}(h)$, $k_c^{a,2 1}(h)$ that correspond to antisymmetric eigenfunctions. The matrix of these eigenvalues can be diagonalized to yield $k_c^{s, \pm}(h)$, $k_c^{a, \pm}(h)$. See~\cite{Peng:2017kro,Peng:2017spg} for more computational details. 
\end{enumerate}
Notice that the 4-point functions receiving contributions from the diagonal kernels do not mix with those receiving contributions from the non-diagonal kernels. Therefore in the following, we call the 4-point functions constructed by these kernels the ``diagonal" and ``non-diagonal" channels respectively. 

The different eigenvalues obey the following relationship~\cite{Peng:2017spg}:
\bal
k_c^{a,d}(h) = k_c^{s,-}(h + \frac{1}{2}) = k_c^{a,+}(h - \frac{1}{2})  \\
k_c^{s,d}(h) = k_c^{s,+}(h - \frac{1}{2}) = k_c^{a,-}(h + \frac{1}{2})\,,
\eal
which is a manifestation of supersymmetry among the operators propagating in the different channels.

The spectrum of physical operators that run in this set of 4-point function are determined by the condition that at least one of the eigenvalues equals to one. The dimensions of the operators in the different channels are determined by the $h$ that satisfy 
\bal\label{eq:diveigenvalues0}
k_c^{a/s,\pm}(h) = 1\ .
\eal
For example the lightest multiplets are found to satisfy the equations
\bal
\label{eq:diveigenvalues}
k_c^{a,d}(\frac{3}{2}) = k_c^{s,-}(2) = k_c^{a,+}(1) = 1  \\
k_c^{s,d}(\frac{3}{2}) = k_c^{s,+}(1) = k_c^{a,-}(2) = 1\,,
\eal
where we find two multiplets each consisting of 1 spin-1, 2 spin-$\frac{3}{2}$ and 1 spin-2 operators.

The sum of ladder diagrams can be evaluated as a geometric sum of the diagonalized kernels, which is schematically of the form $\frac{1}{1 - K}$ acting on the zero-rung basis. One can further decompose it into a sum over a complete set of orthonormal eigenfunction basis that diagonalize the kernels with $h=\frac{1}{2}+i \, s$, which constitutes the principle series, as well as $h\in \mathbb{Z}$, which constitutes the discrete set. However, one need to check if any of the eigenvalues corresponding to these eigenfunctions is one; when this happens the above geometric sum diverges.   
This in general is not a problem since the solution to the eigenvalue equation~\eqref{eq:diveigenvalues0} are mostly irrational. 
However, the aforementioned supermultiplets consisting  $h=1,3/2,2$ operators do appear in the set of orthonormal eigenfunctions and the eigenvalues of the kernels acting on them give 1. So they lead to genuine divergences of 4-point functions in the conformal limit~\cite{Fu:2016vas, Peng:2017spg}, like the fermionic model~\cite{Maldacena:2016hyu}. 

This is simply a signature that such operators actually corresponds to zero modes in the space of solutions to the Schwinger-Dyson equation in the conformal limit: their presence is due to the spontaneous breaking of the supersymmetric reparameterization symmetry of the conformal limit of the Schwinger-Dyson equation. To regularize this divergence, one has to also introduce a small explicit  breaking of the conformal symmetry, which amounts to correct the eigenvalues of equation~(\ref{eq:diveigenvalues}) by stepping outside of the conformal limit. The simplest approach to do so is to consider the large-$q$ limit where we can solve the model without relying on the conformal symmetry.

\section{Green's functions}
\label{sec:largeqpropagators}
We start by finding the exact propagators of the supersymmetric SYK model in the large-$q$ limit.

\subsection{The $\mathcal{N}=1$ model}
We consider the following ansatz for fermionic and bosonic large-$q$ propagators. 
\bal
G_{\psi}(\t) &= \frac{1}{2} \text{sgn}(\t) \big( 1 + \frac{1}{q} g_{\psi}(\t) + \cdots \big)  \\
G^{b}(\t) &= - \delta(\t) + \frac{1}{2 q} g_{b}(\t) + \cdots\,, \label{largeq}
\eal
where $g_\psi(-\t)=g_\psi(\t)$ and  $g_b(-\t)=g_b(\t)$. Notice that this is slightly different from (2.34) of the \cite{Fu:2016vas}, but we will nevertheless show that it reproduces their result.  
The propagators in the frequency domain read
\bal
G_{\psi}(\omega) &= - \frac{1}{i \omega} + \frac{1}{2 q} (\text{sgn} \circ \tilde{g}_{\psi}(\omega))  \\
G_b(\omega) &= - 1 + \frac{1}{2 q} \tilde{g}_b(\omega)
\eal
where $\circ$ refers to convolutions in frequency space and $\tilde{g}_{\psi,b}$ is the Fourier transform of the $g_{\psi,b}$ functions. They can be inverted and then plugged into the Schwinger-Dyson equations of the propagators. To the first order in $1 / q$, this leads to
\bal
G_{\psi}(\omega)^{-1} &= - i \omega + \frac{\omega^2}{2 q} \text{sgn} \circ \tilde{g}_{\psi}(\omega) = - i \omega - \Sigma_{\psi}(\omega)  \\
G_b(\omega)^{-1} &= -1 - \frac{1}{2 q} \tilde{g}_b(\omega) = -1 - \Sigma_{\psi}(\omega)\ .
\eal
Solving the equations gives the expressions for the self energies, which can further be transformed back to the time domain to get
\bal
\label{eq:schwd}
\Sigma_{\psi}(\t) &= \frac{1}{2 q} \partial_{\t}^2 \big( \text{sgn}(\t) g_{\psi}(\t) \big)  \,,\qquad 
\Sigma_{b}(\t) = \frac{1}{2 q} g_{b}(\t)\ .
\eal
On the other hand, the self energies can also be computed as
\bal
\Sigma_{\psi}(\t) &= (q-1) J G^b(\t) (G^{\bar{\psi}}(\t))^{q-2}  \,, \qquad 
\Sigma_{b}(\t) = J (G^{\psi}(\t)^{q-1}\ .
\eal
Plugging in the large-$q$ ansatz for propagators~\eqref{largeq}, we get
\bal
\Sigma_{\psi}(\t) &= \frac{q-1}{q} \frac{J}{2^{q-1}} \text{sgn}(\t) g_b(\t) e^{g_{\psi}(\t)}  \,, \qquad
\Sigma_{b}(\t) = \frac{J}{2^{q-1}} e^{g_{\psi}(\t)} 
\eal
In the large-$q$ limit, we keep $\mathcal{J} = q J / 2^{q-2}$ fixed, similar to large-$q$ analysis of other analogous models~\cite{Maldacena:2016hyu,Fu:2016vas}.  Rewriting the previous set of equations in terms of $\mathcal{J}$, and 
comparing with equation~(\ref{eq:schwd}), we get
\bal
\mathcal{J}^2 \text{sgn}(\t) e^{2 g_{\psi}(\t)} &= \partial_{\t}^2 \big( \text{sgn}(\t) g_{\psi}(\t) \big)  \,, \qquad
\mathcal{J} e^{g_{\psi}(\t)} = g_{b}(\t)\ .
\eal
They agree with (2.35) of~\cite{Fu:2016vas} (up to factors of sgn$(\t)$ on both sides of the $g_{\psi}$ equation, the effect of this factor is not observed in this computation that is constrained in $[0,\beta)$.) After rescaling $g_{\psi}$ by a factor of 2, the $g_{\psi}$ equation is identical to (2.16) of \cite{Maldacena:2016hyu}, and it has a solution~\cite{Maldacena:2016hyu}
\bal
\label{eq:gpsisol}
e^{2 g_{\psi}(\t)} &= \cos ^2\left(\frac{\pi  v}{2}\right) \sec ^2\left(\pi  v \left(\frac{1}{2}-\frac{|\t|}{\b}\right)\right)  \\
\b \mathcal{J} &= \frac{\pi v}{\cos( \pi v / 2)}\ .
\eal
Next we consider a similar computation for the $\mathcal{N}=2$ model. 

\subsection{The $\mathcal{N}=2$ model}

In the $\mathcal{N} = 2$ model, we assume the following large - $q$ propagators,
\bal
G_{\psi}(\t) &= \frac{1}{2} \text{sgn}(\t) \big( 1 + \frac{1}{q} g_{\psi}(\t) + \cdots \big) \,,\quad 
G_{\bar{\psi}}(\t) = \frac{1}{2} \text{sgn}(\t) \big( 1 + \frac{1}{q} g_{\bar{\psi}}(\t) + \cdots \big) \label{Gpsi} \\
G_{b}(\t) &= - \delta(\t) + \frac{1}{2 q} g_{b}(\t) + \cdots \,,~\quad\qquad 
G_{\bar{b}}(\t) = - \delta(\t) + \frac{1}{2 q} g_{\bar{b}}(\t) + \cdots \,,\label{Gb}
\eal
where $G_{O}(\t)=\langle\mathcal{T} O(\t)\bar{O}(0)\rangle$. Inverting them and plugging them into the Schwinger-Dyson equation leads to the following expressions for the self energies
\bal
\Sigma_{b}(\t) &= \frac{1}{2 q} g_{b}(\t) \,,\qquad\qquad\qquad
\Sigma_{\bar{b}}(\t) = \frac{1}{2 q} g_{\bar{b}}(\t)\,,  \label{sigmab}\\
\Sigma_{\psi}(\t) &= \frac{1}{2 q} \partial_{\t}^2 \big( \text{sgn}(\t) g_{\psi}(\t) \big)\,, \quad
\Sigma_{\bar{\psi}}(\t) = \frac{1}{2 q} \partial_{\t}^2 \big( \text{sgn}(\t) g_{\bar{\psi}}(\t) \big)\ .\label{N2self}
\eal
On the other hand, self energies in the $\mathcal{N} = 2$ model are defined as
\bal
\Sigma^{b}(\t) &= J (G^{\psi}(\t)^{q-1} \,, \qquad\qquad \quad\quad~~~ 
\Sigma^{\bar{b}}(\t) = J (G^{\bar{\psi}}(\t)^{q-1}\,, \label{eq:selfens1} \\
\Sigma^{\psi}(\t) &= (q-1) J G^{b}(\t) (G^{\bar{\psi}}(\t))^{q-2} \,,\quad 
\Sigma^{\bar{\psi}}(\t) = (q-1) J G^{\bar{b}}(\t) (G^{\psi}(\t))^{q-2}\ . \label{eq:selfens2}
\eal
After substituting the expressions of propagators and comparing with the other expressions of the self energies~\eqref{eq:selfens1} and~\eqref{eq:selfens2}, we get
the following set of equations,
\bal
g_{b}(\t) &= \mathcal{J} e^{g_{\psi}(\t)}\,,\qquad \partial_{\t}^2 \big( \text{sgn}(\t) g_{\psi}(\t) \big) =  \mathcal{J} \text{sgn}(\t) g_b(\t) e^{g_{\bar{\psi}}(\t)}  \\
g_{\bar{b}}(\t) &= \mathcal{J} e^{g_{\bar{\psi}}(\t)} \,,\qquad \partial_{\t}^2 \big( \text{sgn}(\t) g_{\bar{\psi}}(\t) \big) = \mathcal{J} \text{sgn}(\t) g_{\bar{b}}(\t) e^{g_{\psi}(\t)}
\eal
Eliminating $g_b(\t)$ and $g_{\bar{b}}(\t)$ from the equations, we get
\bal
\partial_{\t}^2 \big( \text{sgn}(\t) g_{\psi}(\t) \big) =  \mathcal{J}^2 \text{sgn}(\t) e^{g_{\psi}(\t) g_{\bar{\psi}}(\t)} =
\partial_{\t}^2 \big( \text{sgn}(\t) g_{\bar{\psi}}(\t) \big)\ .
\eal
We thus find that our ansatz~\eqref{Gpsi} and~\eqref{Gb} lead to identical $g_{\psi}(\t)$ and $g_{\bar{\psi}}(\t)$ up to a linear function in $\t$. Finiteness of the Green's functions at large time forbids such linear terms and hence we conclude that $g_{\psi}(\t) = g_{\bar{\psi}}(\t)$, $g_{b}(\t) = g_{\bar{b}}(\t)$. We then conclude that they are both solved by equation~(\ref{eq:gpsisol}) which we recast here
\bal
e^{2 g_{\psi}(\t)} &= \cos ^2\left(\frac{\pi  v}{2}\right) \sec ^2\left(\pi  v \left(\frac{1}{2}-\frac{|\t|}{\b}\right)\right)  \\
\b \mathcal{J} &= \frac{\pi v}{\cos( \pi v / 2)}\ .\label{eq:gpsisolN2}
\eal
Notice that the $v\to 1$ limit is equivalent to the $\b \mathcal{J} \to \infty$ limit.

\section{The regularized 4-point functions}
\label{sec:largeq}
Next we consider the diverging contribution to the 4-point function, which is regularized by slightly stepping away from the conformal limit. We discuss the ``non-diagonal" and ``diagonal" kernels respectively. 

\subsection{The non-diagonal channel}
\label{sec:nond}
Kernels of equation~(\ref{eq:nondiagonal}) act on a two-component vector of ``eigenfunctions" 
\bal
\label{eq:defs}
K^{1 1} \psi_{1} = k_{1 1} \psi_{1} \,\,,\,\,
K^{1 2} \psi_{2} = k_{1 2} \psi_{1} \,\,,\,\,
K^{2 1} \psi_{1} = k_{2 1} \psi_{2}\,,
\eal
whose repeated action can be conveniently encoded  into repeated multiplication by the following matrix 
\bal
\begin{pmatrix}
 \langle \psi_{1}|   K^{11}|\psi_{1}\rangle & \langle\psi_{1}| K^{12}|\psi_{2}\rangle \\
  \langle \psi_{2}| K^{21}|\psi_{1}\rangle & 0 \\
\end{pmatrix} 
~
\begin{pmatrix}
k_{11} & k_{12} \\
k_{21} & 0 \\
\end{pmatrix} 
\eal
where $k^{ij}$ are the numbers in~\eqref{eq:defs}. We can first diagonalize this matrix of $k^{ij}$ and then the computation of the geometric series of the matrix can be trivialized. 

Concretely, the eigenequations read
\bal
k_{11} \psi_1(\t_1,\t_2) &=J q^2 \int d \t_3 d \t_4 \frac{\text{sgn}(\t_{14})}{2}
\frac{\text{sgn}(\t_{2 3})}{2}\\
&\qquad\quad \times
\left( -\delta(\t_{3 4}) + \frac{1}{2 q} g_{b}(\t_{3 4}) \right) 
\frac{\text{sgn}(\t_{3 4})^{q-3}}{2^{q-3}}
e^{g_{\psi}(\t_{3 4})}
\psi_{1}(\t_3, \t_4)  \\
k_{1 2} \psi_1(\t_1,\t_2) &=q J \int d \t_3 d \t_4 
\frac{\text{sgn}(\t_{1 4})}{2}
\frac{\text{sgn}(\t_{2 3})}{2}
\frac{\text{sgn}(\t_{3 4})^{q-2}}{2^{q-2}}
e^{g_{\psi}(\t_{3 4})} \psi_2(\t_3,\t_4)  \\
k_{2 1} \psi_2(\t_1,\t_2)&=q J \int d \t_3 d \t_4 \delta(\t_{1 4}) \delta(\t_{2 3})
\frac{\text{sgn}(\t_{3 4})^{q-2}}{2^{q-2}}
e^{g_{\psi}(\t_{3 4})} \psi_{1}(\t_3,\t_4)\ . 
\eal
Those equations can be recast into differential form after applying $\partial_{\t_1} \partial_{\t_2}$ to both sides,
\bal
\label{eq:diffeqs}
\mathcal{J}^2 e^{2 g_{\psi}(\t_{2 1})}
\psi_{1}(\t_2,\t_1) &=
k_{1 1} \frac{\partial}{\partial \t_1}
\frac{\partial}{\partial \t_2}
\psi_1(\t_1,\t_2)
 \\
\mathcal{J} \text{sgn}(\t_{2 1}) 
e^{g_{\psi}(\t_{21})}
\psi_2(\t_2,\t_1) &=
k_{1 2} 
\frac{\partial}{\partial_{\t_1}}
\frac{\partial}{\partial_{\t_2}} \psi_{1}(\t_1,\t_2)  \label{eq:diffeqs2}\\
\mathcal{J} \text{sgn}(\t_{2 1})
e^{g_{\psi}(\t_{2 1})}
\psi_{1}(\t_2,\t_1) &=
k_{2 1} \psi_{2}(\t_1,\t_2)\ .\label{eq:diffeqs3}
\eal
Eliminating the $\psi_1(\t_1,\t_2)$ in the equations, we obtain
\bal
\label{eq:diffbosonic1}
\mathcal{J}^2 \sgn(\t_{1 2}) e^{g_{\psi}(\t_{1 2})} \psi_2(\t_1,\t_2) = 
k_{1 2} k_{2 1} \frac{\partial}{\partial \t_1} \frac{\partial}{\partial \t_2} \sgn(\t_{2 1})
e^{- g_{\psi}(\t_{1 2})}
\psi_2(\t_1,\t_2)
\eal
Following~\cite{Maldacena:2016hyu}, we take the following Fourier ansatz 
\bal
\label{eq:diffbosonic2}
&\psi_2(\t_1,\t_2) = \frac{e^{- i  n y}}{\sin(\tilde{x}/2)} \psi_{2,n}(x)
\,\,\,\, , \,\,\,\,
\tilde{x} = v x + (1 - v) \pi  \\
&\left( n^2 + 4 \partial_x^2 - \frac{v^2}{k_{12} k_{2 1} \sin^2(\tilde{x} / 2)} \right)
\psi_{2,n}(x) = 0\,,\label{eigeneq}
\eal
where
\bal
x=\t_1-\t_2\,, \qquad y =\frac{\t_1+\t_2}{2}\ .
\eal
Notice that we did not assume any symmetry properties of $\psi_1$ and $\psi_2$ in deriving equation~(\ref{eq:diffbosonic1}) and (\ref{eq:diffbosonic2}). The general solution of equation~(\ref{eq:diffbosonic2}) is
\bal
\psi_{2,h,n}(x)&=c_1 \psi_{2,h,n}^{(1)}(x)+c_2 \psi_{2,h,n}^{(2)}(x)\\\label{sol2n}
\psi_{2,h,n}^{(1)}(x) &= (\sin \frac{\tilde{x}}{2})^h \,_2F_1\left(\frac{h - \tilde{n}}{2}, \frac{h + \tilde{n}}{2}; \frac{1}{2}; \cos^2\left(\frac{\tilde{x}}{2}\right)\right)  \\
\psi_{2,h,n}^{(2)}(x) &=
\cos \frac{\tilde{x}}{2}
(\sin \frac{\tilde{x}}{2})^h
\,_2F_1\left(\frac{1 + h - \tilde{n}}{2}, \frac{1 + h + \tilde{n}}{2}; \frac{3}{2}; \cos^2\left(\frac{\tilde{x}}{2}\right)\right)\,,
\eal
where $c_1$, $c_2$ are arbitrary constants and $\tilde{n} = n / v$. In the above expressions, we have also set $\b = 2\pi$  and identified $ k_{1 2} k_{2 1}  = \frac{1}{h (h-1)}$ after which the eigenfunction with dimension $h$ solves the equation~\eqref{eigeneq}.
The solution $\psi_{1,n}$ can be obtained by substituting this solution into the third equation of equation~(\ref{eq:diffeqs}). Then the first equation of equation~(\ref{eq:diffeqs}) can be solved as follows. Rewriting this equation in terms of $\psi_2(\t)$, we get (after making use of $e^{g_{\psi}(\t_1,\t_2)} = e^{g_{\psi}(\t_2,\t_1)}$),
\bal
\label{eq:consistency}
\mathcal{J}^2 \sgn(\t_{1 2}) e^{g_{\psi}(\t_{2 1})} \psi_2(\t_1,\t_2) = 
k_{1 1} \frac{\partial}{\partial \t_1} \frac{\partial}{\partial \t_2} \sgn(\t_{2 1})
e^{- g_{\psi}(\t_{1 2})}
\psi_2(\t_2,\t_1)\ .
\eal
Next, we eliminate the RHS of~\eqref{eq:diffeqs} with the help of~\eqref{eq:diffeqs2} and rewrite the LHS of ~\eqref{eq:diffeqs} with the help of~\eqref{eq:diffeqs3}, we get
\bal
k_{2 1} k_{12}\psi_{2}(\t_1,\t_2) &=
k_{11} 
\psi_2(\t_2,\t_1)\ .
\eal
It is then useful to decompose $\psi_2$ into the symmetric and antisymmetric basis in $\t_1$ and $\t_2$: $\psi_2(\t_1,\t_2) = \psi_2^S(\t_1,\t_2) + \psi_2^A(\t_1,\t_2)$, where $\psi_2^S(\t_1,\t_2)=\psi_2^S(\t_2,\t_1)$ and  $\psi_2^A(\t_1,\t_2)=-\psi_2^A(\t_2,\t_1)$. Then we get
\bal
&k_{1 2} k_{2 1} 
(\psi_2^A(\t_1,\t_2) + \psi_2^S(\t_1,\t_2))
 =  k_{1 1}
(\psi_2^S(\t_1,\t_2) - \psi_2^A(\t_1,\t_2))
\eal
which can be rearranged into the form
\bal
(k_{1 1} - k_{1 2} k_{2 1}) 
\psi_2^S(\t_1,\t_2)
= 
(k_{1 1} + k_{1 2} k_{2 1}) 
\psi_2^A(\t_1,\t_2)
\eal
The nontrivial solutions of this equation are $k_{1 2} k_{2 1} + k_{1 1} = 0 = \psi_2^S(\t_1,\t_2)$, which means $\psi_2$ is antisymmetric and $\psi_1$ symmetric; or $k_{1 2} k_{2 1} - k_{1 1} =0 = \psi_2^A(\t_1,\t_2)$, which means $\psi_2$ is symmetric an $\psi_1$ is antisymmetric.

\subsubsection{The conformal Limit}
\label{sec:cl}

The eigenvalues of the non-diagonal kernels in the 
conformal limit were found in \cite{Fu:2016vas,Peng:2017spg} 
\bal
k_c^{a,\pm}(h) &= \mp \frac{\Gamma(2 - \frac{1}{q}) \Gamma(1 - \frac{h}{2} - \frac{1}{2 q}) \Gamma(\frac{1}{2 q} + \frac{h}{2}) \Gamma(\frac{1}{2} - h + \frac{1}{q} \mp \frac{1}{2})}{\Gamma(1 + \frac{1}{q}) \Gamma(1 + \frac{h}{2} - \frac{1}{2 q}) \Gamma(\frac{1}{2 q} - \frac{h}{2}) \Gamma(\frac{3}{2} - h - \frac{1}{q} \mp \frac{1}{2})}  \\
k_c^{s,\pm}(h) &= \mp \frac{\Gamma(2 - \frac{1}{q}) \Gamma(\frac{1}{2} - \frac{h}{2} - \frac{1}{2 q}) \Gamma(\frac{1}{2} + \frac{h}{2} + \frac{1}{2 q}) \Gamma(\frac{1}{2} - h + \frac{1}{q} \mp \frac{1}{2})}{\Gamma(1 + \frac{1}{q}) \Gamma(\frac{1}{2} + \frac{h}{2} - \frac{1}{2 q}) \Gamma(\frac{1}{2} + \frac{h}{2} - \frac{1}{2 q}) \Gamma(\frac{3}{2} - h - \frac{1}{q} \mp \frac{1}{2})}
\eal
In the limit of $q \rightarrow \infty$, they become
\bal
\label{eq:confq}
k_c^{a,+}(h) = -\frac{1}{h} \,,\qquad 
k_c^{a,-}(h) = \frac{1}{h-1}  \,,\qquad 
k_c^{s,+}(h) = \frac{1}{h} \,,\qquad 
k_c^{s,-}(h) = \frac{1}{1 - h}
\eal
Now we try to reproduce these results from our large $q$ analysis. Diagonalizing the matrix
\bal
\begin{pmatrix}
    k_{11} & k_{12} \\
    k_{21} & 0 \\
\end{pmatrix}
\eal
with $k_{11}$, $k_{12}$, $k_{21}$ defined by (\ref{eq:defs}) leads to 
\bal
\tilde{k}^{\pm} = \frac{k_{1 1} \pm \sqrt{k_{11}^2 + 4 k_{1 2} k_{2 1}}}{2}
\eal
Next recall that our large $q$ analysis leads to $k^s_{1 1} = -k_{1 2} k_{2 1}$ and $k^a_{1 1} =  k_{1 2} k_{2 1}$, where $k^s_{11}$ corresponds to the case of even $\psi_1(x)$ and $k^a_{11}$ corresponds to the case of odd $\psi_1(x)$. So we expect the following identification
\bal
\label{eq:kaspm}
\tilde{k}^{a,\pm}&\sim  \frac{k^a_{1 1} \pm \sqrt{\left(k^a_{11}\right)^2 + 4 k_{1 2} k_{2 1}}}{2} = \frac{k_{1 2} k_{2 1} \pm \sqrt{(k_{1 2} k_{2 1})^2 + 4 k_{1 2} k_{2 1}}}{2}  \\
\tilde{k}^{s,\pm}&\sim  \frac{k^s_{1 1} \pm \sqrt{\left(k^s_{11}\right)^2 + 4 k_{1 2} k_{2 1}}}{2} = \frac{- k_{1 2} k_{2 1} \pm \sqrt{(k_{1 2} k_{2 1})^2 + 4 k_{1 2} k_{2 1}}}{2}\ .
\eal
In the conformal limit $v = 1$, (\ref{eq:diffbosonic2}) tells us that the conformal weight $h$ of $\psi_2$ is given by $k_{1 2} k_{21} = \frac{1}{h (h-1)}$. Substituting this into (\ref{eq:kaspm}) leads to
\bal
\label{eq:eigs+-}
\tilde{k}^{a,\pm} &= \frac{1}{2 h (h - 1)} \pm \sqrt{\Big( \frac{1 - 2h}{2 h ( h - 1 )} \Big) ^2 } = \{ -\frac{1}{h} , \frac{1}{h - 1} \}  \\
\tilde{k}^{s,\pm} &= \frac{1}{2 h (1 - h)} \pm \sqrt{\Big( \frac{1 - 2h}{2 h ( h - 1 )} \Big) ^2 } = \{ \frac{1}{h} , \frac{1}{1 - h} \}\,,
\eal
which matches the results from (\ref{eq:confq}). Note that there are only two eigenvalues in the strict $q\to \infty$ limit leading to operators with non-negative dimension: $k(h) = 1$: $k^{a,-}(2) = 1$ and $k^{s,+}(1) = 1$. According to \cite{Peng:2017spg}, they live in the same supermultiplet. Furthermore, their conformal dimension lives on the integration contour of the 4-point function, so they lead to actual divergence of the 4-point functions. To regulate such divergences, we go slightly away from the conformal limit. This is discussed in the next section.

\subsubsection{Stepping out of the conformal limit}\label{sectionSA}
Next we consider the contribution to the 4-point function from the $h=2$ supermultiplet, which correspond to the super-reparameterization modes. In the large-$q$ limit we can explicitly regulate their contributions by stepping away from the conformal limit. 

We first analyze the symmetry property of the eigenfunctions in our problem. We first consider the eigenfunction $\psi_2$ whose two external legs, which are to be fused to the kernels, are both bosonic and are periodic
\bal
\psi_2(\t_1 + 2 \pi, \t_2) = \psi_2(\t_1,\t_2) \,\,\, , \,\,\,
\psi_2(\t_1, \t_2 + 2 \pi) = \psi_2(\t_1,\t_2)\ .
\eal
Notice that here and in the following of this section we set $\b=2\p$ for simplicity. 
The swap statistics further leads to
\bal
\psi_2^S (\t_1, \t_2) = \psi_2^S (\t_2,\t_1) \,,\qquad \psi_2^A (\t_1, \t_2) =- \psi_2^A (\t_2,\t_1)\,,
\eal
where the two different symmetries under the swapping of the two fields are both possible because the two fields at the two positions are conjugate to each other, instead of being the same. As a result, we get
\bal
\psi_2^S(2\pi-x,y+\pi)=\psi_2^S(x,y)\,,\qquad \psi_2^A(2\pi-x,y+\pi)=-\psi_2^A(x,y)\ ,
\eal
We then check the property of our solution under $(x,y)\to (2\pi-x,y+\pi)$. Using the fact that the transformation $x\to 2\pi -x$ translates to $\tilde{x} = v x + (1 - v)\pi \to -v x + (1 + v)\pi = 2\pi -\tilde{x} $, the symmetry property of the solution~\eqref{sol2n} around $x = \pi$ is manifest
\bal
\psi_2^{(1)}(2\pi-x)=\psi_2^{(1)}(x)\,,\qquad \psi_2^{(2)}(2\pi-x)=-\psi_2^{(2)}(x)\ . 
\eal
It is then clear that the symmetric and antisymmetric pieces of the solution read 
\bal
\psi_{2,h}(x)&= \psi_{h}^{S}(x)+ \psi_{h}^{A}(x)\label{solSA}\\
\psi_{h}^{S}(x) &= c_1 \sum_{n= 2\mathbb{Z}} \frac{e^{- i  n y}}{\sin(\tilde{x}/2)} \psi_{2,h,n}^{(1)}(x) + c_2 \sum_{n= 2\mathbb{Z}+1} \frac{e^{- i  n y}}{\sin(\tilde{x}/2)} \psi_{2,h,n}^{(2)}(x)\label{egfS}\\
\psi_{h}^{A}(x) &=c_1 \sum_{n= 2\mathbb{Z}+1} \frac{e^{- i  n y}}{\sin(\tilde{x}/2)} \psi_{2,h,n}^{(1)}(x) + c_2 \sum_{n= 2\mathbb{Z}} \frac{e^{- i  n y}}{\sin(\tilde{x}/2)} \psi_{2,h,n}^{(2)}(x)\,,\label{egfA}
\eal
where we have written out the explicit $h$ dependence in the eigenfunctions for the convenience of the later discussion.
It is not surprising that such a rewriting is possible since in the non-diagonal channels, the fields on the ladder-rungs are all bosonic and hence one should be able to rewrite eigenfunctions in the set of basis of the bosonic eigenfunctions. However, a crucial difference  from the results in~\cite{Maldacena:2016hyu} is that in our expressions the $n$ can be both even and odd for either of the expressions. This doubling precisely correspond to the fact that there are two $h=2$ multiplets, one corresponds to the super-reparameterization mode and the other one is expected to be a normal operator. Comparing with the results in~\cite{Maldacena:2016hyu}, we find $\psi_2^S(x)$ is precisely the solution considered there for the SYK model based on Majorana fermions and it is clear that it correspond to the super-reparameterization mode. The $\psi^A_2(x)$ corresponds to the other spin-2 super-multiplet.  

We first try to understand better the nature of the second $h=2$ multiplet by checking whether it leads to a divergence in the 4-point function and hence correspond to zero modes.  The way we verify this is to consider the process of going to the conformal limit, which is controlled by the limit $v\to 1$. 
Explicitly, we first consider the divergence of the eigenfunctions in the conformal limit $v=1$, we get
\bal
 \frac{e^{- i  n y}}{\sin(\tilde{x}/2)} \psi_{2,h,n}^{(1)}(x) &
\text{ diverges at } h=1\,, n=2\mathbb{Z}\text{ and } h=2\,,  n=2\mathbb{Z}+1\cup \{0\}\,,\label{div1}\\
\frac{e^{- i  n y}}{\sin(\tilde{x}/2)} \psi_{2,h,n}^{(2)}(x) &
\text{ diverges at } h=1\,, n=2\mathbb{Z}+1\cup \{0\}  \text{ and } h=2\,,  n=2\mathbb{Z}\cup \{\pm 1\} \ .\label{div2}
\eal 
First recall that the $n=0,\pm 1$ modes at $h=2$ and the $n=0$ mode at $h=1$ of the above expansion correspond to the global $sl(2,\mathbb{R})$ symmetry of the solution. They are true zero modes of the model that we want to remove. Therefore, in the following discussion, we will not consider them since they are always removed from the theory. 

Next notice that when we consider the $v\to 1$ limit, the eigenfunctions are expected to approach the eigenfunction in the conformal limit, in particular the mode near $h=2$ are expected to approach the $h=2$ mode in the conformal limit that leads to the divergence of the 4-point functions. Notice that this divergence is the property of the associated eigenvalues and the eigenfunctions themselves should be well defined and normalizable. From the above analysis, in particular~\eqref{egfA}, \eqref{egfS}, \eqref{div1} and \eqref{div2} we see precisely that the $\psi^S_{1}$ and $\psi^A_{2}$ diverges in the limit $v \to 1$. This means that in the conformal limit the would-be eigenfunctions associated to the antsymmetric spin-2 operator and the symmetric spin-1 operator all diverge.  Since we only consider normalizable  eigenfunctions of the kernel when computating the 4-point functions, we then conclude that the operators associated to the would-be $\psi^S_{1}$ and $\psi^A_{2}$ eigenfunctions in the conformal limit actually decouple from the theory in the conformal limit, and they do not enter the correlation functions. This is compatible with the fact that these two operators are in the same $h=1$ multiplet. There decoupling means there is only one spin-2 operator in the conformal limit of the theory, which is simply the stress tensor.

Given this we consider only the other two components, namely $\psi^S_{2}$ and $\psi^A_{1}$, that correspond to the stress tensor and a spin-1 operator in the same multiplet in the conformal limit. Although these eigenfunctions do not diverge in the strict conformal limit $v\to 1$, they do lead to divegence of the correlation function due to the divergence of the geometric sum of the kernel. We thus seek to go away from the conformal limit so that the dimension is away from the conformal answer $h=2,1$. Hence the kernel no longer evaluates to 1 and the divergence is regulated. The process of going away from the conformal limit can be conveniently parameterized by the values of $v$ that are different from 1. As $v$ increases from 1, the dimension $h$ of the real eigenfunctions get corrected so that the eigenfunctions remains  normalizable as the theory is driven away from the conformal limit. To determine such corrections to the dimension $h$, we can simply requre the eigenfunctions $\psi^S_{2}$ and $\psi^A_{1}$ to be regular at any $v$ around 1. 

The condition for this to happen is similar to~\cite{Maldacena:2016hyu}, namely one of the first two arguments of the eigenfunction should vanish. 
This means the dimension should have the following dependence on $v$
\bal
\psi^A_{1}: \qquad &h_1(v) = 1 + \tilde{n} - n = 1 + n \frac{1 - v}{v}\\
\psi^S_{2}: \qquad &h_2(v) = 2 + \tilde{n} - n = 2 + n \frac{1 - v}{v} 
\,,
\eal
that renders both $\psi^S_{2}$ and $\psi^A_{1}$ to be finite at $v\neq 1$. Here we have used $h_1(v)$ to represent the dimension of the spin-1 operator corresponding to the $\psi^A_{1}$ eigenfunction and $h_2(v)$ to represent the operator correspond to the $\psi^S_{2}$ eigenfunction away from the conformal limit. Given this correction, it is simple to determine what are the eigenvalues of the non-diagonal kernel away from the conformal limit at any value $v$ 
\bal
k^{s,+} &= \frac{1}{h_1(v)} = \frac{1}{1 + n (1 - v)/v}\label{eq:seigen}
\\
k^{a,-} &= \frac{1}{h_2(v) - 1} = 
\frac{1}{1 + n (1 - v)/v}
\label{eq:aeigen}
 \ .
\eal
Since these eigenvalues of the kernels are now away from 1, the corresponding geometric series converges and the 4-point function is thus well defined and sensible. Furthermore, note that the seeming mismatch of the $S,A$ label on the eigenfunction $\psi$ and that on the eigenvalue $k^{a/s,\pm}$ is not a typo, this is simply because the eigenfunctions we considered here are the second component in the vector that has the opposite symmetry property as the (first component of the) multiplet.

\subsection{The diagonal channel}
\label{sec:appdiagonal}

\begin{figure}
	\centering
	\includegraphics[width=0.8\textwidth]{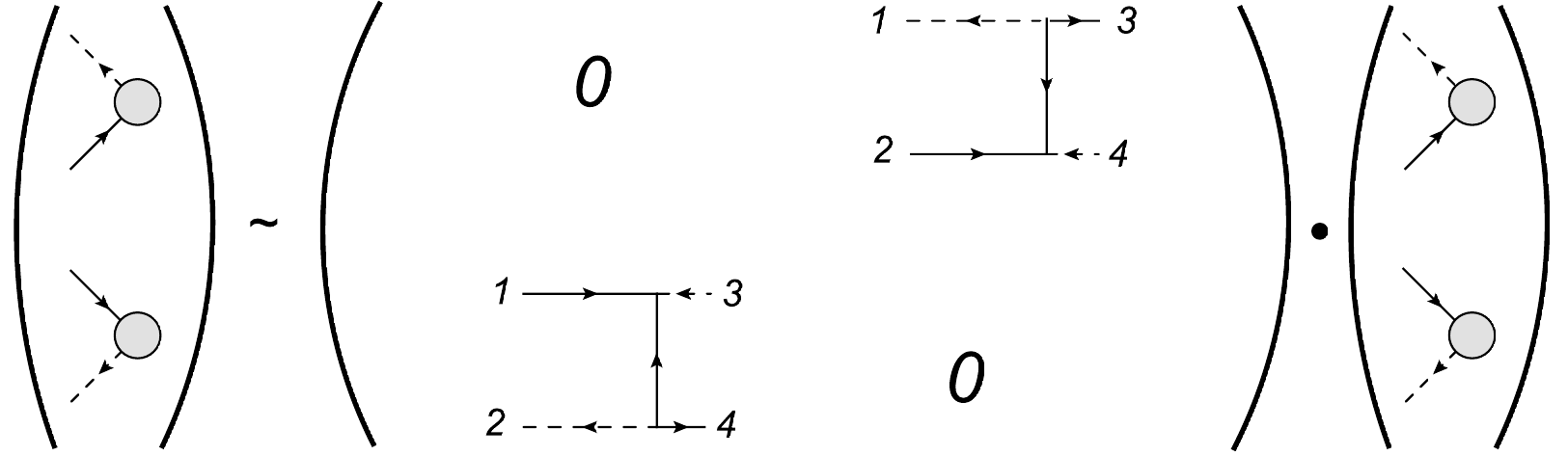}
	\caption{Action of the diagonal Kernels $K_d^{(1)}, K_d^{(2)}$. In the figure the real lines represent fermionic propagators and the dashed lines represent bosonic propagators.}\label{fig:diagonalk}
\end{figure}

The diagonal kernels act on a two-component vector of eigenfunctions as
\bal
K_d^{(1)} \psi_2 = k_1 \psi_1\,,\qquad K_d^{(2)} \psi_1 = k_2 \psi_2\,,
\eal
where the $\psi_i$ depend on two times: in $\psi_1(\t_1,\t_2)$ the line attached to $\t_1$ is bosonic and the line attached to $\t_2$ is fermionic, in $\psi_2(\t_1,\t_2)$ the line attached to $\t_1$ is fermionic and the line attached to $\t_2$ is bosonic, as illustrated in figure~\ref{fig:diagonalk}.

Their repeated actions can again be represented by repeated multiplication of the matrix 
\bal
\label{eq:matrixdiagonal}
\begin{pmatrix}
    0 & \langle \psi_1 |K_d^{(1)}|\psi_2 \rangle \\
    \langle \psi_2 |K_d^{(2)}|\psi_1 \rangle  & 0 \\
\end{pmatrix} 
~=~\begin{pmatrix}
    0 & k_1 \\
    k_2 & 0\\
\end{pmatrix} \ .
\eal
 The kernels $K_d^{(1)}$, $K_d^{(2)}$ are
\bal
K_{d}^{(1)} &= J (q - 1) G^{\psi}(\t_{1 3}) G^{b}(\t_{2 4}) \big( G^{\psi}(\t_{3 4}) \big)^{(q - 2)}  \\
K_{d}^{(2)} &= J (q - 1) G^{b}(\t_{1 3}) G^{{\psi}}(\t_{2 4}) \big( G^{\psi}(\t_{3 4}) \big)^{(q - 2)}\ .
\eal

Working to the leading order in $1 / q$ and plugging in the expression~\eqref{Gpsi}, \eqref{Gb}, \eqref{sigmab} and ~\eqref{N2self}, 
the eigen-equation for $K_d^{(1)}$ is thus
\bal
- \frac{J (q-1)}{2^{q-1}} \int d \t_3 d \t_4 \text{sgn}(\t_{1 3}) \delta(\t_{2 4}) \text{sgn}(\t_{3 4}) e^{g_{\psi}(\t_{3 4})} \psi_2(\t_3 , \t_4) = k_1 \psi_1(\t_1, \t_2)
\eal
Applying $\partial_{\t_1}$ to both sides and integrating over $\t_3$ and $\t_4$, we get
\bal
\label{eq:diag21}
- \mathcal{J} \text{sgn}(\t_{12}) e^{g_{\psi}(\t_{12})} \psi_1(\t_1,\t_2) = k_1 \partial_{\t_1} \psi_2(\t_1,\t_2)\ .
\eal
Repeating the previous calculation for $K_d^{(2)}$ yields
\bal
\label{eq:diag12}
 \mathcal{J} \text{sgn}(\t_{12}) e^{g_{\psi}(\t_{12})} \psi_2(\t_1,\t_2) = k_2 \partial_{\t_2} \psi_1(\t_1,\t_2)\ .
\eal
Substituting (\ref{eq:diag21}) into (\ref{eq:diag12}), we obtain
\bal
\mathcal{J}^2 \text{sgn}(\t_{21}) e^{g_{\psi(\t_{2 1})}} \psi_2(\t_1,\t_2) =  -k_1 k_2 \frac{\partial}{\partial \t_2} \left( \text{sgn}(\t_{21}) e^{-g_{\psi}(\t_{2 1})} \frac{\partial}{\partial \t_1} \psi_2(\t_1,\t_2)   \right)\ .
\eal
Assuming the ansatz 
\bal
	&\psi_2(\t_1,\t_2) = \frac{e^{- i  n y}}{\sin^{\frac{1}{2}}(\tilde{x}/2)} \psi^d_{2,n}(x)\,,\qquad 
	\tilde{x} = v x + (1 - v) \pi\,,
\eal
where $n$ is now half-integer since the eigenfunction is fermionic, the above equation has a solution
\bal
    \psi^d_{2,n}(x) &= e^{-\frac{1}{2} \left(i \tilde{x}\right)} (-1)^{-\frac{n}{v}} \left(-1+e^{i \tilde{x}}\right)^{\frac{1}{k}+\frac{1}{2}} \left(e^{i \tilde{x}}\right)^{-\frac{2 n+v}{4 v}} \\
    &\quad \times \left(c_1 (-1)^{n/v} \left(e^{i \tilde{x}}\right)^{\frac{n}{v}+\frac{1}{2}} \, _2F_1\left(\frac{1}{k},\frac{n}{v}+\frac{1}{2}+\frac{1}{k};\frac{n}{v}+\frac{1}{2};e^{i \tilde{x}}\right)\right.\\
    &\qquad\qquad  \left.+i c_2 e^{i \tilde{x}} \, _2F_1\left(1+\frac{1}{k},-\frac{n}{v}+\frac{1}{2}+\frac{1}{k};\frac{3}{2}-\frac{n}{v};e^{i \tilde{x}}\right)\right)
\eal
where  $k^2 = k_1 k_2$ is the square of the eigenvalues of the matrix~\eqref{eq:matrixdiagonal}. The $h=\frac{3}{2}$ mode, which is in the same supermultiplet that contains the spin-1 and
 spin-2 modes corresponding to $\psi_1^A$ and $\psi_2^S$, is an eigenfunction of this kernel matrix~\eqref{eq:matrixdiagonal} with eigenvalue $k=1$ and therefore could potentially lead to divergence. As in the previous nondiagonal case, we first require that this eigenfunction remains  finite in the $v\to 1$, $k\to 1$, $x\to 0$ limit for any $|n|>\frac{1}{2}$.\footnote{Here the $k\to 1$ limit is equivalent to $h= \frac{3}{2}$.} This condition removes half of the solution by setting
 \bal
 c_1 =0 \,, \qquad \forall\, n>\frac{1}{2}\,,\qquad \text{ and } \qquad 
 c_2 =0 \,, \qquad \forall\, n<-\frac{1}{2}\ .
 \eal
 The remaining half modes are true eigenvalues that leads to a divergence of the 4-point functions $\langle \psi^i b^i \bar\psi^j \bar{b}^j \rangle$. To regulate such divergences, we slightly move away from the conformal limit, namely we consider $v \neq 1$. As the previous non-diagonal case, we determine the eigenvalues $k$ of the matrix~\eqref{eq:matrixdiagonal} at $v\neq 1$ by requiring the eigenfunctions to be finite around $x = 0$. This can only be true when either of the first two arguments is a non-positive integer or half integer. For example, when $n>0$ the term proportional to $c_2$ contributes and we require
\bal
	-\frac{n}{v}+\frac{1}{2}+\frac{1}{k}= n_0\,,\label{knp}
\eal   
where $n_0$ is a non-positive integer or half integer. 
This determines
\bal
	n_0=\frac{3-2n}{2}\,,
\eal 
and the solution of the equation~\eqref{knp} is 
\bal
k=\frac{1}{1-n (1-1/v)} \,, \qquad \forall\, n>\frac{1}{2}\ .
\eal
Similarly, for negative modes, we require
\bal
\frac{n}{v}+\frac{1}{2}+\frac{1}{k}= \frac{3+2n}{2}\,,\label{knn}
\eal 
which leads to   
\bal
k=\frac{1}{1+n (1-1/v)} \,, \qquad \forall\, n<-\frac{1}{2}\ .
\eal
Combining the two cases we get the expression of the eigenvalues of the diagonal kernel~\eqref{eq:matrixdiagonal} away from the conformal limit
\bal
    k = \frac{1}{1 - |n| (1-1/v)}\ .\label{oddshift}
\eal
The shift of the eigenvalue is proportional to $|n|$, which is similar to the results in the nondiagonal kernels.

With all the ingredients above, we can compute the regularized contribution from the soft-modes to the 4-point function, which is just geometric sums of the kernels on the eigenfunctions of the $h=2$ multiplets where the eigenvalues are shifted as in~\eqref{eq:seigen},~\eqref{eq:aeigen}   and~\eqref{oddshift}. The details of this computation is in exact parallel with the computation in~\cite{Peng2017a} so we do not repeat here.

\section{Chaos exponents}
\label{sec:chaos}
We can compute the chaotic behavior of the supersymmetric model in the large-$q$ limit as well. Since we expect that the largest exponent is again due to the spin-2 reparameterization mode, we only focus on the non-diagonal channels. 

We compute the chaotic exponent by diagonalizing the set of retarded kernels following~\cite{KitaevTalk1,KitaevTalk2, Maldacena:2016hyu,Peng:2017spg}. The retarded kernels are defined to be  
\bal
\label{eq:nondiagonalchaos}
K^{1 1}_R &= J \frac{(q-1)!}{(q-3)!} G_R^{\psi}(\t_{1 4}) G_R^{\bar{\psi}}(\t_{2 3}) G_{lr}^{\bar{b}}(\t_{3 4}) \big( G_{lr}^{\psi}(\t_{3 4}) \big)^{q-3}  \\
K^{1 2}_R &= J \frac{(q-1)!}{(q-2)!} G_R^{\psi}(\t_{1 4}) G_R^{\bar{\psi}}(\t_{2 3}) \big( G_{lr}^{\psi}(\t_{3 4}) \big)^{q-2} \\
K^{2 1}_R &= J \frac{(q-1)!}{(q-2)!} G_{R}^{\bar{b}}(\t_{1 4}) G_R^{b}(\t_{2 3}) \big( G_{lr}^{\psi}(\t_{3 4}) \big)^{q-2}\,,
\eal
where the retarded propagators $G_R$ and the Wightman propagators $G_{lr}$ are obtained from the Euclidean propagators by analytically continuation.  

To the leading order of the large-$q$ limit, the retarded propagators are given by $G_R^{\psi}(t) = \q(t)$ and $G_b(t) = - \delta(t)$ which satisfy the SUSY relation $G_b(t) = - \partial_t G_{\psi}(t)$. We also need to compute the large-$q$ expression of the left-right blob in the ladder kernel, for this we need to analytically continue large powers of fermionic propagators
\bal
\Sigma^\psi(\t) \sim    G^{\psi}(\t)^{q-2} = \frac{\sgn(\t)^{q-2}}{2^{q-2}} e^{g_{\psi}(\t)}\,,
\eal
Because in the supersymmetric model the $q$ is odd, the continuation to real time clearly depends on from where we do the continuation. For example, in the $\t > 0$ region, the continuation of $G^{\psi}(\t)^{q-2} = \frac{1}{2^{q-2}} e^{g_{\psi}}(\t)$ leads to 
\bal
 \frac{1}{2^{q-2}} e^{g_{\psi}(\b / 2 + i t)} = \frac{1}{2^{q-2}} \frac{\cos(\pi v / 2)}{\cosh(\pi v t)} \equiv \frac{1}{2^{q-2}} e^{\tilde{g}_{\psi}(t)}\ .
\eal
In the $\t < 0$ region,  $G^{\psi}(\t)^{q-2} = - \frac{1}{2^{q-2}} e^{g_{\psi}}(\t)$ we get
\bal
     \frac{1}{2^{q-2}} e^{g_{\psi}(\b / 2 + i t)} = - \frac{1}{2^{q-2}} \frac{\cos(\pi v / 2)}{\cosh(\pi v t)} = - \frac{1}{2^{q-2}} e^{\tilde{g}_{\psi}(t)}\ .
\eal
On the other hand, the continuation to the left-right form of the product $G^{\psi}(\t)^{q-3} G^b(\t) = \frac{1}{2^{q-2}} \frac{1}{q} g_b(\t) e^{g_{\psi}(\t)}$ does not depend on the sign of $\t$ due to the even power $q-3$: the continuation simply gives
\bal
    \frac{1}{2^{q-3}} \frac{1}{q} \mathcal{J} e^{2 g_{\psi}(\b/2+it)} = \frac{1}{2^{q-2}} \frac{1}{q} \mathcal{J} \frac{\cos(\pi v / 2)^2}{\cosh(\pi v t)^2}\ .
\eal
To compute the Lyapunov exponent in the large - $q$ limit, we diagonalize the retarded kernels with the above large-$q$ expressions and an exponentially growing ansatz. The eigenequations are similar to the Euclidean computation~\eqref{eq:defs} 
\bal
    K^{11}_R \psi_1 = k^R_{11} \psi_1 \,,\qquad  K^{1 2}_R \psi_2 = k^R_{1 2} \psi_1 \,,\qquad K^{2 1}_R \psi_1 = k^R_{2 1} \psi_2\,,
\eal
where $(\psi_1 \,, \psi_2)$ is a two-component vector of eigenfunctions.

In terms of the large-$q$ expressions, they become
\bal
    & \mathcal{J}^2\int d t_3 d t_4 \q(t_{1 4}) \q(t_{2 3}) e^{2 \tilde{g}_{\psi}(t_{3 4})} \psi_{1} (t_3, t_4) = k_{1 1} \psi_1(t_1, t_2)  \\
    &\mathcal{J} \int d t_3 d t_4 
    \q(t_{1 4})
    \q(t_{2 3})
    \sgn(t_{3 4})
    e^{\tilde{g}_{\psi}(t_{3 4})}
    \psi_2(t_3,t_4) 
    = k_{1 2} \psi_1(t_1,t_2)  \\
    &\mathcal{J} \int d t_3 d t_4 \delta(t_{1 4}) \delta(t_{2 3})
    sgn(t_{3 4})
    e^{\tilde{g}_{\psi}(t_{3 4})}
     \psi_{1}(t_3,t_4) 
    = k_{2 1} \psi_2(t_1,t_2)\ .
\eal
Applying $\partial_{t_1} \partial_{t_2}$ on both sides of the first two equations, and integrating over the delta functions, we get
\bal
\label{eq:diffeqqs}
&\mathcal{J}^2 e^{2 \tilde{g}_{\psi}(t_{2 1})}
\psi_{1}(t_2,t_1) = k_{1 1}
 \frac{\partial}{\partial t_1}
\frac{\partial}{\partial t_2}
\psi_1(t_1,t_2)
 \\
&\mathcal{J} \sgn(t_{2 1}) 
e^{\tilde{g}_{\psi}(t_{21})}
\psi_2(t_2,t_1) = k_{1 2}
\frac{\partial}{\partial t_1}
\frac{\partial}{\partial t_2} \psi_{1}(t_1,t_2)  \\
&\mathcal{J} \sgn(t_{2 1})
e^{\tilde{g}_{\psi}(t_{2 1})}
\psi_{1}(t_2,t_1) = k_{2 1}
\psi_{2}(t_1,t_2)\ .
\eal
Plugging the expression for $\psi_2$ from the third equation into the second one, we obtain
\bal
    \mathcal{J}^2 \sgn(t_{1 2}) e^{\tilde{g}_{\psi}(t_{1 2})} \psi_2(t_1,t_2) = k_{1 2} k_{2 1}
    \frac{\partial}{\partial t_1} \frac{\partial}{\partial t_2} \sgn(t_{2 1}) e^{- \tilde{g}_{\psi}(t_{1 2})} \psi_2(t_1,t_2)\ .
\eal
Assuming an ansatz of the form $\psi_2(t_1,t_2) = \frac{e^{\lambda_L (t_1 + t_2) / 2}}{\cosh(v (t_1-t_2) / 2)} u(t_{1}-t_{2})$, the above equation simplifies to a form 
\bal
    \left( \frac{\lambda_L^2}{4} -  \frac{d^2}{d x^2} \right) u(x) = \frac{\pi^2 v^2}{\b^2 k_{1 2} k_{2 1}} \frac{1}{\cosh^2(\pi v x / \b)} u(x)  \,,
\eal
where $x=t_1-t_2 $. In terms of $\tilde{x} = \pi v x / \b$, this equation becomes
\bal
\label{eq:cosh2potential}
    \left( \frac{\lambda_L^2 \b^2}{4 \pi^2 v^2} - \frac{d^2}{d \tilde{x}^2} \right) \tilde{u}(\tilde{x}) = \frac{1}{k_{1 2} k_{2 1}} \frac{1}{\cosh^2(\tilde{x})} \tilde{u}(\tilde{x})\ .
\eal
The physical value of the Lyapunov exponent $\lambda_L$ renders at least one of the eigenvalues to be 1. Given the expression of the matrix of the eigenvalues of the retarded kernels
\bal
\begin{pmatrix}
k^R_{11} & k^R_{12} \\
k^R_{21} & 0 \\
\end{pmatrix} \,,
\eal
at least one eigenvalue equal to 1 means the eigenvalue equation
\bal
	\mu(\mu-k^R_{11})-k_{12}^R k^R_{21}=0\,,
\eal
has a solution at $\mu=1$. This means $k^R_{1 1} + k^R_{1 2} k^R_{2 1} = 1$. Similar to the analysis in section~(\ref{sec:nond}), we must have either $k^R_{1 1} = k^R_{1 2} k^R_{2 1}$, corresponding to $\psi_1$ being antisymmetric and $\psi_2$ symmetric, or $k^R_{1 1} = - k^R_{1 2} k^R_{2 1}$, corresponding to $\psi_1$ being symmetric and $\psi_2$ antisymmetric. According to the analysis in section~\ref{sectionSA}, only the multiplet with a symmetry property of the first case is present in the spectrum. This leads to $k^R_{1 2} k^R_{2 1} = k^R_{1 1} = \frac{1}{2}$. At this value the  equation~(\ref{eq:cosh2potential}) is recognized as the Schr\"{o}dinger equation discribing a particle moving in a $V(\tilde{x}) = -2 / \cosh^2 (\tilde{x})$ potential, and the energy is parameterized by $E=-\frac{\lambda_L^2 \b^2}{4 \pi^2 v^2}$. There is one bound state in this potential with energy $E= -1$. This means the Lyapunov exponent is given by
\bal
    \lambda_L = \frac{2 \pi}{\b} v\,,
\eal
which saturates the chaos bound~\cite{Maldacena2016a}.

%%%%%%%%%%%%
%%%%%%%%%%%%%%%%
\section{Effective action}
\label{sec:schwarzian}
%%%%%%%%%%%%%%%%
%%%%%%%%%%%%
Up to now we have regularized the contribution to the 4-point functions from the stress tensor multiplet. The regularized result controlled by the $v\to 1$ limit that, as indicated in~\eqref{eq:gpsisolN2}, controls the leading $\b J$ piece of the 4-point functions. As in the original fermionic SYK model, choosing one solution of the reparameterization invariant infrared Schwinger-Dyson equation breaks the super-reparameterization symmetry spontaneously and our regularization further breaks the super-reparameterization symmetry explicitly. Therefore we expect the would-be Goldstone modes to have finite action. We can write down such an effective action by requiring that it reproduces the leading $\b J$ piece of the 4-point functions. In this section, we derive this effective action explicitly.

We start by performing the disorder average of the random coupling of the action~\eqref{action} and obtain a bilocal action
\bal
S =& -  \log \det( \partial_{\t} - \tilde{\Sigma}_{\bar{\psi} \psi}) +  \log \det \left(\tilde{\Sigma}_{\bar{b} \psi} ( \partial_{\t} - \tilde{\Sigma}_{\bar{\psi} \psi})^{-1} \tilde{\Sigma}_{\bar{\psi} b} - \delta(\t) - \tilde{\Sigma}_{\bar{b} b} \right)
\\
&-  \log \det( \partial_{\t} - \tilde{\Sigma}_{\psi \bar{\psi}}) +  \log \det \left(\tilde{\Sigma}_{b  \bar{\psi}} ( \partial_{\t} - \tilde{\Sigma}_{\psi \bar{\psi}})^{-1} \tilde{\Sigma}_{\psi \bar{b}} - \delta(\t) - \tilde{\Sigma}_{b \bar{b}} \right)  \\
&+ \int d \t_1 d \t_2 \Big( \tilde{\Sigma}_{\psi \bar{\psi}}(\t_1,\t_2) \tilde{G}_{\psi \bar{\psi}}(\t_1,\t_2) +\tilde{\Sigma}_{ \bar{\psi}\psi}(\t_1,\t_2) \tilde{G}_{\bar{\psi}\psi }(\t_1,\t_2) \\
&+
\tilde{\Sigma}_{b\bar{b} }(\t_1,\t_2) \tilde{G}_{b\bar{b} }(\t_1,\t_2) \tilde{\Sigma}_{\bar{b} b}(\t_1,\t_2) \tilde{G}_{\bar{b} b}(\t_1,\t_2) 
+ \tilde{\Sigma}_{\bar{b} \psi}(\t_1,\t_2) \tilde{G}_{\bar{b} \psi}(\t_1,\t_2)\\
& + \tilde{\Sigma}_{b \bar{\psi}}(\t_1,\t_2) \tilde{G}_{b \bar{\psi}}(\t_1,\t_2) 
- J \tilde{G}_{\bar{b} b}(\t_1,\t_2) \tilde{G}_{\bar{\psi} \psi}(\t_1,\t_2)^{q-1} 
 \\
&- J \tilde{G}_{b \bar{b}}(\t_1,\t_2) \tilde{G}_{\psi \bar{\psi}}(\t_1,\t_2)^{q-1} - J (q-1) \tilde{G}_{\bar{b} \psi}(\t_1, \t_2) \tilde{G}_{\bar{\psi} b}(\t_1,\t_2) \tilde{G}_{\bar{\psi} \psi}(\t_1,\t_2)^{q-2}\\
&
- J (q-1) \tilde{G}_{b \bar{\psi}}(\t_1, \t_2) \tilde{G}_{\psi \bar{b}}(\t_1,\t_2) \tilde{G}_{\psi \bar{\psi}}(\t_1,\t_2)^{q-2} 
\Big) \,,
\eal
where we have inserted the following  Lagrange multiplier constraints
\bal
\label{eq:lagrangem}
    &\int \mathcal{D} \tilde{G}_{\bar{\psi} \psi} \mathcal{D} \tilde{\Sigma}_{\bar{\psi} \psi} \exp \left( - N \tilde{\Sigma}_{\bar{\psi} \psi}(\t_1,\t_2) (\tilde{G}_{\bar{\psi} \psi}(\t_1,\t_2) - \frac{1}{N} \bar{\psi}^i(\t_1) \psi^i(\t_2))  \right) = 1
     \\
    &\int \mathcal{D} \tilde{G}_{ \psi \bar{\psi}} \mathcal{D} \tilde{\Sigma}_{ \psi\bar{\psi}} \exp \left( - N \tilde{\Sigma}_{ \psi \bar{\psi}}(\t_1,\t_2) (\tilde{G}_{ \psi\bar{\psi}}(\t_1,\t_2) - \frac{1}{N} {\psi}^i(\t_1) \bar{\psi}^i(\t_2))  \right) = 1
    \\
    &\int \mathcal{D} \tilde{G}_{\bar{b} b} \mathcal{D} \tilde{\Sigma}_{\bar{b} b} \exp \left( - N \tilde{\Sigma}_{\bar{b} b}(\t_1,\t_2) (\tilde{G}_{\bar{b} b}(\t_1,\t_2) - \frac{1}{N} \bar{b}^i(\t_1) b^i(\t_2))  \right) = 1
    \\
    &\int \mathcal{D} \tilde{G}_{ b \bar{b}} \mathcal{D} \tilde{\Sigma}_{ b \bar{b}} \exp \left( - N \tilde{\Sigma}_{b \bar{b}}(\t_1,\t_2) (\tilde{G}_{b\bar{b} }(\t_1,\t_2) - \frac{1}{N} {b}^i(\t_1) \bar{b}^i(\t_2))  \right) = 1
     \\
     &\int \mathcal{D} \tilde{G}_{\bar{b} \psi} \mathcal{D} \tilde{\Sigma}_{\bar{b} \psi} \exp \left( - N \tilde{\Sigma}_{\bar{b} \psi}(\t_1,\t_2) (\tilde{G}_{\bar{b} \psi}(\t_1,\t_2) - \frac{1}{N} \bar{b}^i(\t_1) \psi^i(\t_2))  \right) = 1\\
     &\int \mathcal{D} \tilde{G}_{{b} \bar{\psi}} \mathcal{D} \tilde{\Sigma}_{{b} \bar{\psi}} \exp \left( - N \tilde{\Sigma}_{{b} \bar{\psi}}(\t_1,\t_2) (\tilde{G}_{{b} \bar{\psi}}(\t_1,\t_2) - \frac{1}{N} {b}^i(\t_1) \bar{\psi}^i(\t_2))  \right) = 1
     \\
     &\int \mathcal{D} \tilde{G}_{\bar{\psi} b} \mathcal{D} \tilde{\Sigma}_{\bar{\psi} b} \exp \left( - N \tilde{\Sigma}_{\bar{\psi} b}(\t_1,\t_2) (\tilde{G}_{\bar{\psi} b}(\t_1,\t_2) - \frac{1}{N} \bar{\psi}^i(\t_1) b^i(\t_2))  \right) = 1
      \\
     &\int \mathcal{D} \tilde{G}_{{\psi} \bar{b}} \mathcal{D} \tilde{\Sigma}_{{\psi} \bar{b}} \exp \left( - N \tilde{\Sigma}_{{\psi} \bar{b}}(\t_1,\t_2) (\tilde{G}_{{\psi} \bar{b}}(\t_1,\t_2) - \frac{1}{N} {\psi}^i(\t_1) \bar{b}^i(\t_2))  \right) = 1\ .
\eal
Further notice that because the time dependence are different the above equations of the quantity $G_{\bar{O}O}$ and $G_{O\bar{O}}$ are independent.
A fermion number conserving solution to the set of saddle point equations, which is simply the set of Schwinger-Dyson equations, of the  $\tilde{G}$, $\tilde\Sigma$ fields from the above action consists of the following vanishing components
\bal
     G_{\bar{b} \psi} = G_{\bar{\psi} b} = G_{b \bar{\psi}} = G_{\psi \bar{b}} = 0
    \,,\qquad 
    \Sigma_{\bar{b} \psi} = \Sigma_{\bar{\psi} b} = \Sigma_{b \bar{\psi}} = \Sigma_{\psi \bar{b}} = 0\,,
\eal
as well as the other nonvanishing components that satisfy
\bal
	&( \partial_{\t} - \Sigma_{\psi \bar{\psi}})* G_{\bar{\psi}\psi } = \delta
\,,\,\qquad
	( \partial_{\t} - \Sigma_{\bar{\psi} \psi})* G_{ \psi \bar{\psi}} = \delta \,,
	\\
	&(- \delta - \Sigma_{b \bar{b}}) *G_{\bar{b}b} = \delta  \,, \qquad\quad  
	(- \delta - \Sigma_{\bar{b} b}) *G_{b\bar{b}} = \delta\,,
\eal
where we have used the untilded $G_{ij}$, $\Sigma_{ij}$ to denote the solutions to the Schwinger-Dyson equations, and we have also used the property $G_{\bar{\psi}\psi}(\t)=-G_{\psi\bar{\psi}}(-\t)$ and $G_{\bar{b}b}(\t)=G_{b\bar{b}}(-\t)$.

The would-be Goldstone modes correspond to the spontaneously broken super-reparameterization symmetry of the Schwinger-Dyson equations, whose action can be realized as infinitesimal deformation of the solution. Notice that here we consider the solution of the full Schwinger-Dyson equation so the super-reparameterization symmetry is also explicitly broken. This is the reason that we will get a finite effective action at the end. Notice that if this symmetry were not explicitly broken, the Goldstone mode should map solutions of the Schwinger-Dyson equations to the solution of the equations. But since in our case this symmetry is explicitly broken from the beginning, we do not require the deformed solution to solve the same set of equations. 

Given this, the effective action for such explicit and spontaneous breaking can be obtained by substituting the perturbed solution 
$\tilde{G}_{\psi \bar{\psi}} = G_{\psi \bar{\psi}} + g_{\psi \bar{\psi}}$, $\tilde{G}_{b \bar{b}} = G_{b \bar{b}} + g_{b \bar{b}}$, $\tilde{\Sigma}_{\psi \bar{\psi}} = \Sigma_{\psi \bar{\psi}} + \sigma_{\psi \bar{\psi}}$, 
$\tilde{\Sigma}_{b \bar{b}} = \Sigma_{b \bar{b}} + \sigma_{b \bar{b}}$ into the action and read out the terms of the perturbation to quadratic order. For example, the kinetic terms come from the functional derivatives and read
\bal
    &\frac{1}{2} \log \det( \partial_{\t} - \Sigma_{\psi \bar{\psi}} - \sigma_{\psi \bar{\psi}})= 
    \frac{1}{2} \textrm{tr} \log( \partial_{\t} - \Sigma_{\psi \bar{\psi}} - \sigma_{\psi \bar{\psi}}) \\
    &\qquad =  \frac{1}{4} \textrm{tr} \left( \left(  \partial_{\t} - \Sigma_{\psi \bar{\psi}} \right)^{-1} \sigma_{\psi \bar{\psi}} \left( \partial_{\t} - \Sigma_{\psi \bar{\psi}} \right)^{-1} \sigma_{\psi \bar{\psi}} \right)  \\
    &\qquad =\frac{1}{4} \int d \t_1 \, d \t_2 \, d \t_3 \, d \t_4 \, G_{\bar\psi {\psi}}(\t_{1 2}) \,\sigma_{\psi \bar{\psi}}(\t_{23}) \,  G_{\bar\psi {\psi}}(\t_{3 4}) \, \sigma_{\psi \bar{\psi}}(\t_{41})\ .
\eal
Similarly, we have
\bal
     \frac{1}{2} \log \textrm{det}(-\delta - \Sigma_{b \bar{b}} -  \sigma_{b \bar{b}}) = 
    \frac{1}{4} \int d \t_1 \, d \t_2 \, d \t_3 \, d \t_4 \,  G_{\bar{b} {b}}(\t_{1 2}) \, \sigma_{b \bar{b}}(\t_{ 23})  \, G_{\bar{b} b }(\t_{3 4}) \,  \sigma_{b \bar{b}}(\t_{4 1})\ .
\eal
The quadratic effective action becomes
\bal
    S_{\text{eff}} =&  - \frac{1}{4} \int d \t_1 \, d \t_2 \, d \t_3 \, d \t_4 \, \sigma_{\psi \bar{\psi}}(\t_{1 2}) \,  G_{ \bar{\psi}\psi}(\t_{1 3}) \, G_{ \bar{\psi}\psi}(\t_{2 4}) \,  \sigma_{\psi \bar{\psi}}(\t_{3 4}) 
      \\
    &- \frac{1}{4} \int d \t_1 \, d \t_2 \, d \t_3 \, d \t_4 \, \sigma_{\bar{\psi} \psi}(\t_{1 2}) \,  G_{\psi\bar{\psi} }(\t_{1 3}) \, G_{\psi\bar{\psi} }(\t_{2 4}) \,  \sigma_{\bar{\psi} \psi}(\t_{3 4}) 
      \\
    &+ \frac{1}{4} \int d \t_1 \, d \t_2 \, d \t_3 \, d \t_4 \, \sigma_{b \bar{b}}(\t_{1 2}) \,  G_{\bar{b} b }(\t_{1 3}) \, G_{\bar{b} b }(\t_{2 4}) \,  \sigma_{b \bar{b}}(\t_{3 4})
      \\
    &+ \frac{1}{4} \int d \t_1 \, d \t_2 \, d \t_3 \, d \t_4 \, \sigma_{\bar{b} b}(\t_{1 2}) \,  G_{ b \bar{b} }(\t_{1 3}) \, G_{\bar{b} b }(\t_{2 4}) \,  \sigma_{\bar{b} b}(\t_{3 4})  \\
    &+ \frac{1}{4} \int d \t_1 \, d \t_2 \, d \t_3 \, d \t_4 \, \sigma_{\bar{b} \psi}(\t_{1 2}) G_{b\bar{b} }(\t_{1 3}) G_{\psi\bar{\psi} }(\t_{2 4}) \sigma_{b \bar{\psi}}(\t_{3 4})  \\
    &+ \frac{1}{4} \int d \t_1 \, d \t_2 \, d \t_3 \, d \t_4 \, \sigma_{b \bar{\psi}}(\t_{1 2}) G_{\bar{b} b }(\t_{1 3}) G_{ \bar{\psi}\psi}(\t_{2 4}) \sigma_{\psi \bar{b}}(\t_{3 4})  \\
     &+ \int d \t_1 d \t_2 \Big( g_{\psi\bar{\psi}} \sigma_{\psi\bar{\psi}} + g_{\bar{\psi}\psi} \sigma_{\bar{\psi}\psi} + g_{\bar{b}b} \sigma_{\bar{b}b} +
     g_{\bar{b}b} \sigma_{\bar{b}b} +
     g_{\bar{b} \psi} \sigma_{\bar{b} \psi} + g_{b \bar{\psi}} \sigma_{b \bar{\psi}}  \\
    &  - J {q-1 \choose 2} G_{b \bar{b}} G_{\psi \bar{\psi}}^{q-3}  g_{\psi \bar{\psi}}^2- J{q-1 \choose 2} G_{\bar{b} b} G_{\bar{\psi} \psi}^{q-3}  g_{\bar{\psi} \psi}^2  - (q-1) J G_{\psi \bar{\psi}}^{q-2} g_{b \bar{b}} g_{\psi \bar{\psi}} \\
    &
    - (q-1) J G_{\bar{\psi} \psi}^{q-2} g_{ \bar{b} b} g_{\bar{\psi} \psi}
    - (q-1) J G_{\bar{\psi} \psi}^{q-2} g_{\bar{b} \psi} g_{\bar{\psi} b} 
    - (q-1) J G_{\psi \bar{\psi}}^{q-2} g_{b \bar{\psi}} g_{\psi \bar{b}} \Big)\ .
\eal
Next we integrate out the $\sigma_{ij}$ fields to get an effective action of the $g_{ij}$  fields
\bal
\label{eq:effact}
    S_{\text{eff}} =& 
        g * \begin{pmatrix}
    X & 0 \\
    0 & Y 
    \end{pmatrix}^{-1} * g +
    \bar{g} * \begin{pmatrix}
    \bar{X} & 0 \\
    0 & \bar{Y} 
    \end{pmatrix}^{-1} * \bar{g}
    + g_{\bar{b} \psi} * Z * g_{\bar{\psi} b} + g_{b \bar{\psi}} * Z * g_{\psi \bar{b}}  \\
    &- \int d \t_1 d \t_2 \Big( J{q-1\choose 2}G_{b \bar{b}} G_{\psi \bar{\psi}}^{q-3}  g_{\psi \bar{\psi}}^2
    + J {q-1\choose 2} G_{\bar{b} b} G_{\bar{\psi} \psi}^{q-3}  g_{\bar{\psi} \psi}^2
    + (q-1) J G_{\psi \bar{\psi}}^{q-2} g_{b \bar{b}} g_{\psi \bar{\psi}}  \\
    &- (q-1) J G_{\bar{\psi} \psi}^{q-2} g_{\bar{b} b} g_{\bar{\psi} \psi}
    + (q-1) J G_{\bar{\psi} \psi}^{q-2} g_{\bar{b} \psi} g_{\bar{\psi} b}
    + (q-1) J G_{\psi \bar{\psi}}^{q-2} g_{b \bar{\psi}} g_{\psi \bar{b}} 
    \Big)\,,
\eal
where ``$*$" represents convolution and $g=(g_{\psi\bar{\psi}},g_{b\bar{b}})$, $\bar{g}=(g_{\bar\psi{\psi}},g_{\bar{b}{b}})$. The $X$, $Y$, $Z$ actions are respectively
\bal
    (X * f)(\t_1,\t_2) &= \int d \t_3 \, d \t_4 G_{\bar{\psi}\psi}(\t_{1 3}) G_{\bar{\psi}\psi}(\t_{2 4})  f(\t_3, \t_4) \,, \\
    (Y *  f)(\t_1,\t_2)& = \int d \t_3 \, d \t_4 G_{\bar{b} b}(\t_{1 3}) G_{\bar{b} b}(\t_{2 4}) f(\t_{3},\t_4)  \,,\\
    (Z *  f)(\t_1,\t_2) &= \int d \t_3 \, d \t_4 G_{\bar{\psi}\psi }(\t_{2 4}) G_{\bar{b}b} (\t_{1 3}) f(\t_3,\t_4)\ .
\eal
We can then rewrite equation~(\ref{eq:effact}) as
\bal
    S_{\text{eff}} =& 
    g * \begin{pmatrix}
    X & 0 \\
    0 & Y 
    \end{pmatrix}^{-1} * g
    + \bar{g} * \begin{pmatrix}
    \bar{X} & 0 \\
    0 & \bar{Y} 
    \end{pmatrix}^{-1} * \bar{g}  \\
  &  + g_{b \bar{\psi}} * Z * g_{\bar{b} \psi} + g_{b \bar{\psi}} * \bar{Z} * g_{\psi \bar{b}}
    - \int d \t_1 d \t_2 \Bigg( g  \begin{pmatrix}
    U & V \\ V & 0
    \end{pmatrix}  g + \bar{g}   \begin{pmatrix}
    \bar{U} & \bar{V} \\ \bar{V} & 0
    \end{pmatrix}  \bar{g}  \\
& + (q-1) J G_{\bar{\psi} \psi}^{q-2} g_{\bar{b} \psi} g_{\bar{\psi} b}
    + (q-1) J G_{\psi \bar{\psi}}^{q-2} g_{b \bar{\psi}} g_{\psi \bar{b}} 
    \Bigg)\,,
\eal
where $U = J \frac{(q-1)(q-2)}{2} G_b G_{\psi}^{q-3}$, $V = J \frac{q-1}{2} G_{\psi}^{q-2}$, and we have absorbed all the integrals in to the convolution notation ``$*$". A further change of variables to 
\bal
    \tilde{g} = \sqrt{\begin{pmatrix}
    U & V \\ V & 0
    \end{pmatrix}} g \,\,\,,\,\,\,
    \tilde{g}_{\bar{b} \psi} = \sqrt{(q-1)J} G_{\psi}^{(q-2)/2} g_{\bar{b} \psi} \,\,\,,\,\,\,
    \tilde{g}_{b \bar{\psi}} = \sqrt{(q-1)J} G_{\psi}^{(q-2)/2} g_{b \bar{\psi}}\,,
\eal
puts the action into the form
\bal
    S_{eff} = \tilde{g} * (\tilde{K}^{-1}-\mathbb{I}) * \tilde{g} + \tilde{g}_{\bar{b} \psi} * (\tilde{K}_d^{-1} - \mathbb{I}) * \tilde{g}_{\bar{\psi} b} + \textrm{conjugate}  \,,\label{seff}
\eal
where $\mathbb{I}$ is the identity matrix for convolution, i.e. $\delta(\t_1-\t_3)\delta(\t_2-\t_4)$, and the symmetrized versions of non-diagonal and diagronal kernels are respectively
\bal
\label{eq:finalaction}
    \tilde{K} = \sqrt{\begin{pmatrix}
    U & V \\ V & 0
    \end{pmatrix}} \begin{pmatrix}
    \hat{X} & 0 \\
    0 & \hat{Y} 
    \end{pmatrix} 
    \sqrt{\begin{pmatrix}
    U & V \\ V & 0
    \end{pmatrix}} \,\,\,,\,\,\, 
    \tilde{K}_d = J(q-1)G_{\psi}^{(q-2)/2} \hat{Z} G_{\psi}^{(q-2)/2}\,,
\eal
which precisely agree with the symmetrized version of~\eqref{eq:diagonalk} and~\eqref{eq:nondiagonal} in the sense that the factor of $\sqrt{\begin{pmatrix}
	U & V \\ V & 0
	\end{pmatrix}}$ should be understood as taking ``half" of the left-right blob in the ladder rung, which can be checked as
\bal
\tilde{K}^2=\sqrt{\begin{pmatrix}
	U & V \\ V & 0
	\end{pmatrix}} \begin{pmatrix}
\hat{X} & 0 \\
0 & \hat{Y} 
\end{pmatrix} 
\begin{pmatrix}
	U & V \\ V & 0
\end{pmatrix}
\begin{pmatrix}
\hat{X} & 0 \\
0 & \hat{Y} 
\end{pmatrix} 
\sqrt{\begin{pmatrix}
U & V \\ V & 0
\end{pmatrix}}\,,
\eal  
which contains one more ladder rung than $\tilde{K}$ that is represented by
\bal
K=	\begin{pmatrix}
\hat{X} & 0 \\
0 & \hat{Y} 
\end{pmatrix}
	\begin{pmatrix}
	U & V \\ V & 0
	\end{pmatrix}
 \ .
\eal
Since the symmetric kernels are conjugate to the kernels~\eqref{eq:nondiagonal} 
\bal
\tilde{K}=	R K
R^{-1}
\,,\qquad \qquad R=\sqrt{\begin{pmatrix}
U & V \\ V & 0
\end{pmatrix}}\,,
\eal 
the eigenvector $\tilde{h}$ of the symmetric kernels is related to the eigenvector $h$ of the original kernels as
\bal
\tilde{h}=R h\,,
\eal
where
\bal
K* h = k h\,,\qquad \tilde{K}* \tilde{h}=k \tilde{h}\ .
\eal
Therefore when the ${g}$ in~\eqref{seff} is the (broken) reparameterization of the conformal propagators, the $\tilde{g}$ is an eigenvector of the symmetric kernels with a shifted eigenvalue away from the conformal eigenvalue 1.

In fact, it is easy to check that the variation of the conformal 2-point functions is an eigenfunction of the non-diagonal kernels with eigenvalue one. To show this, we start by varying Schwinger-Dyson equations in the conformal limit,
\bal
\delta_{\e} G_{\psi} \ast \Sigma_{\psi} + G_{\psi} \ast \delta_{\e} \Sigma_{\psi} = 0  \\
\delta_{\e} G_{b} \ast \Sigma_{b} + G_{b} \ast \delta_{\e} \Sigma_{b}   = 0   \,,
\eal
where $f \ast g \equiv \int dt_2 f(t_1,t_2) g(t_2,t_3)$.
We then rewrite the equations using the conformal Schwinger-Dyson equations to
\bal
\label{eq:varied}
- \delta_{\e} G_{\psi} + G_{\psi} \ast \delta_{\e} \Sigma_{\psi} \ast G_{\psi} = 0   \\
- \delta_{\e} G_{b} + G_{b} \ast \delta_{\e} \Sigma_{b} \ast G_{b}  = 0   \ .
\eal
The variations of the self energies are
\bal
\delta_{\e} \Sigma_{b} &= J (q-1) (G_{\psi})^{q-2} \delta_{\e} G_{\psi}  \\
\delta_{\e} \Sigma_{\psi} &= (q-1) J (G_{\psi})^{q-2} \delta_{\e} G_{b} + (q-1) (q-2) J G_{b} (G_{\psi})^{q-3} \delta_{\e} G_{\psi}\ .
\eal
They can be further written in terms of the kernels of equation~(\ref{eq:nondiagonal})
\bal
&- \delta_{\e} G_{\psi} + K_{1 1} \delta_{\e} G_{\psi} + K_{1 2} \delta_{\e} G_{b} = 0  \\
&- \delta_{\e} G_{b} + K_{21} \delta_{\e} G_{\psi} = 0\ .
\eal
In a matrix form this reads
\bal
\begin{pmatrix}
K_{1 1} & K_{1 2} \\
K_{2 1} & 0 
\end{pmatrix}
\begin{pmatrix}
\delta_{\e} G_{\psi} \\
\delta_{\e} G_{b}
\end{pmatrix} = 
\begin{pmatrix}
\delta_{\e} G_{\psi} \\
\delta_{\e} G_{b}
\end{pmatrix}\,,
\eal
which is a simple generalization of the result of the original SYK model~\cite{Maldacena:2016hyu}. Further notice that the statement is true for any variation around the conformal solution.

%%%%%%%%%%%%%%%%%%%%%%%%%%%%%%%%%%
\subsection{Super-Schwarzian action from correlation functions}
%%%%%%%%%%%%%%%%%%%%%%%%%%%%%%%%%%%%%
The above derivation is valid in general. In this section we would like to get the $\b J$ enhanced contribution in the  effective action, where the deformation $g$ and $\bar{g}$ are the reparametrization of the conformal solutions and the kernels also take up the leading correction beyond the value in the conformal limit. 

The time reparametrizations of the conformal propagators, $g = (\delta_{\e} G_{\psi} \,,\, \delta_{\e} G_b)$,  
can be worked out according to the transformation rule of the two point function of primary operators
\bal
\delta_\e G_c =\left(\Delta \e'(\q_1) +\Delta \e'(\q_1) + \e(\q_1)\partial_{\q_1}+ \e(\q_2)\partial_{\q_2}\right)G_c\ .
\eal 
Assuming $\b = 2\pi
$ and plugging $\e(\q)= \sum_n \e_n e^{-i n \q}$ into the conformal propagators, 
\bal
G_{\psi}(\t)=\frac{b_\psi}{\left(\frac{\b}{\pi}\sin(\frac{\pi\t}{\b})\right)^{2\Delta_\psi}}
\,, \qquad
G_b(\t)=\frac{b_b}{\left(\frac{\b}{\pi}\sin(\frac{\pi\t}{\b})\right)^{2\Delta_b}}\,,
\eal
we get 
\bal
\label{eq:variations}
g_{\bar{\psi}\psi}&=\delta_{\e_n} G_{\bar{\psi}\psi} = i \frac{1}{q} \left(\frac{1}{2}\right)^{1/q} b_{\psi} \sin \left( \frac{x}{2} \right)^{-1/q} e^{- i n y} f_n(x)  \\
g_{\bar{b}b}&=\delta_{\e_n} G_{\bar{b}b} = i \frac{q+1}{q} \left(\frac{1}{2}\right)^{1+1/q} b_b \sin \left(\frac{x}{2} \right)^{-(1+1/q)} e^{- i n y} f_n(x) \\
g_{\bar{\psi}b}&= \delta_{\e_n} G_{\bar{\psi}b}=0 \,, \qquad\qquad 
g_{\bar{b}{\psi}}= \delta_{\e_n} G_{\bar{b}{\psi}}=0\,,
\eal
where $f_n(x) =  n \cos \left(\frac{n x}{2} \right) - \sin \left(\frac{n x}{2} \right) \cot \left(\frac{x}{2} \right) 
$ and in the above expressions, we have used the fact that the value of $g_{\bar\psi b}$ and $g_{\bar b\psi}$ vanish in the conformal limit.
On the other hand, the shift of the eigenvalues away from the conformal limit are, to the order of $1/(\b J)$, given by equation~(\ref{eq:seigen}), equation~(\ref{eq:aeigen}). 
Putting all these factors together, the onshell action of the time reparametrizations modes is 
\bal
\label{eq:epsintegral}
S_{\text{eff},\e} &=\tilde{g} (RK^{-1}R^{-1} -\mathbb{I})R g=\tilde{g} * (R k^{-1}-R) {g}\,   = g * R^2 g \,  (k^{-1}-1)\\
&
=2 \int d \t_1 d \t_2 (\delta_{\e_n} G_{\psi}, \delta_{\e_n} G_b) 
    \begin{pmatrix}
    U & V \\
    V & 0
    \end{pmatrix}
    (\delta_{\e_{-n}} G_{\psi}, \delta_{\e_{-n}} G_b) 
    (k^{-1} - 1)   \\
   & =- \frac{q-1}{q^2} \frac{v-1}{v}\frac{(2\p)^2}{\b^2}  \pi^2 J b_{\psi}^q n^2 (n^2 - 1)  \,,
\eal

We now proceed to compute the effective action of the spontaneously and explicitly broken $U(1)$ gauge symmetry.    
The local $U(1)$ transformations $a =a_n e^{i n t}$ acts on the propagators as
\bal
\delta_a G_c =\left(e^{i \alpha_1 a (\q_1)+i \alpha_2 a (\q_2) } -1\right)G_c\,,
\eal
where $\alpha_i$ are the $U(1)$ charge of the fields in the 2-point function. In particular the fundamental fields have the following charges
\bal
\alpha(\bar\psi)=-\frac{1}{q}\,,\qquad \alpha(\psi)=\frac{1}{q}\,,\qquad
\alpha(\bar{b})=\frac{1-q}{q}\,,\qquad \alpha(b)=\frac{q-1}{q}\ .
\eal
Their absolute values are twice the conformal dimensions, which is consistent with the fact that the supermultiplets we considered are all short. We then find
\bal
\label{eq:avariation}
    \delta_{a_n} G_{\psi}& = - \frac{2}{q} e^{i n y} \sin(n x / 2) G_{\psi}  \\
    \delta_{a_n} G_{b}& = 2 \frac{1-q}{q} e^{i n y} \sin(n x / 2) G_b\ .
\eal
To the leading order, the change in eigenvalues is again given by equation~(\ref{eq:seigen}), equation~(\ref{eq:aeigen}), the effective action is then
\bal
\label{eq:aintegral}
S_{\text{eff},a}  &= 
    2 \int d \t_1 d \t_2 (\delta_{a_n} G_{\psi} , \delta_{a_n} G_b) 
    \begin{pmatrix}
    U & V \\
    V & 0
    \end{pmatrix}
    (\delta_{a_{-n}} G_{\psi}, \delta_{a_{-n}} G_b) (k^{-1} - 1)   \\
&=  - 4 \frac{q-1}{q^2} \frac{v-1}{v}  \pi^2 J b_{\psi}^q n^2\ . 
\eal

Next we consider the fermionic transformations, and for simplicity we consider the chiral and anti chiral supersymmetry transformations simultaneously. The transformation of the Green's functions of the primary $\psi$ and $b$ fields gives the following fermionic variations
\bal
g_{\bar{\psi}b}&= -\left( \frac{\bar\h(\t_2)'}{q}G_{\bar\psi\psi}(\t_{1 2})-\bar\h(\t_1)G_{\bar b b}(\t_{1 2})+\bar\h(\t_2)\pa_{\t_2}G_{\bar\psi\psi}(\t_{1 2})\right) \\
g_{\bar{b}\psi}&=-\left(\frac{\h(\t_1)'}{q}G_{\bar\psi\psi}(\t_{1 2})+\h(\t_2)G_{\bar b b}(\t_{1 2})+\h(\t_1)\pa_{\t_1}G_{\bar\psi\psi}(\t_{1 2}) \right)\ .
\eal
Then to the leading order, we get the following effective action 
\bal
\label{eq:etaintegral}
S_{\text{eff},\e} &=\tilde{g}_{\bar{b} \psi} * (\tilde{K}_d^{-1} - \mathbb{I}) * \tilde{g}_{\bar{\psi} b}   = J (q-1)\int d\t_1 d\t_2 G_{\psi}(\t_1,\t_2)^{q-2} g_{\bar{b} \psi}(\t_1,\t_2)  {g}_{\bar{\psi} b}(\t_1,\t_2)  \,  (k_d^{-1}-1)\\
&=
- 2(4n^2-1)\sgn(n) i\p^2 \frac{  q-1}{q^2}J b_\psi^q\frac{2\p}{\b}|n|\frac{1-v}{v}\bar\h_{-n}\h_n\\
 &=
 2(4n^2-1)(-n) i\p^2 \frac{  q-1}{q^2}J b_\psi^q\frac{2\p}{\b}\frac{v-1}{v}\h_n\bar \h_{-n}\,,
 \eal
where the first factor of $2$ is because of an identical contribution from the conjugate channel. 
Notice that in all the above derivations we have omitted some steps of the integrals, whose details can be found in appendix~\ref{integrals}.

Putting everything together, we arrive at a quadratic-order mode expansion of the effective action
\bal
S_{\text{eff}} = &- \alpha \frac{(2\p)^2}{\b^2}\sum_{n \in Z} n^2 (n^2 - 1) \e_n \e_{-n} - 4\alpha \sum_{n \in Z}  n^2 a_n a_{-n}\\
&\qquad \quad
-\alpha \frac{2\p}{\b}\sum_{n \in Z + 1/2} 2  (4 (-in)^2 + 1)(-in) \h_n \bar{\h}_{-n} 
\eal
where $\alpha_S = \pi^2 \frac{q-1}{q^2} \frac{v-1}{v} J b_{\psi}^q$. 
In the position space, this corresponds to the following action quadratic in $\e$, $\h$, $a$,
\bal
S_{\text{eff}} = - \alpha \frac{(2\p)^2}{\b^2}\int d \t \left( (\e'')^2 - (\e')^2 \right) - 4 \alpha \int d \t (\partial a)^2- 2 \alpha \int d \t (4 \h \bar{\h}''' + \h \bar{\h}') 
\ . \label{schker}
\eal
Notice that from our 4-point function computation we can only determine the quadratic order of the effective action. Interactions of these soft modes can be determined via higher point correlation functions, which is out of the scope of the current paper. 

The result reproduces the result in~\cite{Fu:2016vas, Stanford:2017thb}.\footnote{Notice that in our perturbative analysis we only detect the zero-winding, i.e. $n=0$, sector, see e.g.~\cite{Stanford:2017thb} for more general discussions. }
Comparing the bosonic part of this action with the Schwarzian effective action of the complex SYK model, see e.g.~\cite{Bagrets:2016cdf,Davison:2016ngz,Chaturvedi:2018uov,Gu:2019jub}, the $\mathcal{N}=2$ supersymmetry further fixes the relative coefficients between the contribution from the reparameterization mode and the $U(1)$ mode. This relative coefficient is also crucial in connnection to the microscopic entropy counting of the near extremal black holes, see e.g. the discussion in~\cite{Larsen:2019oll}. 

There are other ways of deriving the supersymmetric Schwarzian action. In the following we discuss another approach and show that its results agree with the results from the above computation.

\subsection{Super-Schwarzian action from supersymmetrization}

One can derive the supersymmetric Schwarzian effective action by supersymmetrizing the bosonic Schwarzian effective action. In~\cite{Fu:2016vas}, the bosonic part of the $\mathcal{N}=2$ super-Schwarzian effective action is computed explicitly, here we provide a detailed derivation of the full $\mathcal{N}=2$ effective action and match with our previous computation from the correlation functions (in particular from the summation over the ladder kernels). Following~\cite{Fu:2016vas}, the supersymmetric Schwarzian derivative action is
\bal
I^{\mathcal{N}=2}=\int  d\t d^2\q  S({\t',\q',\bar{\q}';\t,\q,\bar{\q}})=\int  d\t   S^b({\t',\q',\bar{\q}';\t,\q,\bar{\q}})\,, \label{ssch1}
\eal
where the fermionic integral measure is defined as
\bal
\int d^2\q d\q d\bar{\q} =1\,,
\eal
and
\bal
S(\t',\q',\bar{\q}';\t,\q,\bar{\q})=\frac{\partial_{\t}\bar{D}'}{\bar{D}\bar{\q}'}-\frac{\partial_{\t}{D}'}{{D}{\q}'}-2\frac{\partial_{\t}{\q}' \partial_{\t}{\bar\q}'}{({\bar{D}}{\bar\q}')({D}{\q}')}\,,\label{ssch2}
\eal
is the supersymmetric Schwarzian derivative with the superderivatives  defined by
\bal
D=\partial_\q +\bar{\q} \partial_{\t}\,,\qquad \bar{D}=\partial_{\bar\q}+\q \partial_{\t}\ .
\eal
The $(\t',\q',\bar\q')$ are the super-reparameterized supercoordinates.
The bosonic reparameterization transformation reads 
\bal
B: \qquad \t'=f(\t)\,, \qquad \q'=e^{i a(\t)}\sqrt{\partial_\t f(\t)}\q\,,\qquad  \bar\q'=e^{-i a(\t)}\sqrt{\partial_\t f(\t)}\bar\q\ .
\eal
The chiral fermionic piece of the super-reparameterization is
\bal
F: \qquad \t'=\t+\bar\q(\t) \h(\t)\,,\qquad
\q'=\q+\h(\t+\q(\t)\bar{\q}(\t))\,,\qquad
\bar{\q}'=\bar{\q}\,,
\eal
and the anti-chiral fermionic transformation is
\bal
\bar{F}: \qquad \t'=\t+\q(\t) \bar\h(\t)\,,\qquad
\q'=\q\,,\qquad
\bar{\q}'=\bar{\q}+\bar\h(\t-\q(\t)\bar{\q}(\t))\ .
\eal
To get the super-Schwarzian action, we consider succesive actions of $F$, $\bar{F}$ followed by $B$. 
This gives the  super-Schwarzian action
\bal
I^{\mathcal{N}=2}=&\int d\t \left[\text{Sch}(f(\t),\t) -2 (\partial_{\t}  a(\t))^2 
-4{\h}(\t)\partial_\t^3 \bar\h(\t)+8i\partial_\t a(\t){\h}(\t)\partial_\t^2 \bar\h(\t)\right.\\
&\left.\qquad \qquad -2\left(\text{Sch}(f(\t),\t)  -2 (\partial_{\t}  a(\t))^2 -2i \partial^2_{\t}  a(\t) \right){\h}(\t)\partial_\t \bar\h(\t)\right]\,,\label{sn2}
\eal 
where 
\bal
\text{Sch}(f(\t),\t)=\frac{\partial_\t^3 f(\t)}{\partial_{\t} f(\t)}-\frac{3}{2}\left(\frac{\partial_\t^2 f(\t)}{\partial_{\t} f(\t)}\right)^2\,,
\eal
and we have kept only terms that are quadratic in the fermionic transformation variables $\h$ and $\bar{\h}$. We  have also used integration by parts to simplify the result. 

To compare with the above results derived from the correlation function~\eqref{schker}, we  change variable to
\bal
f(\t)=\tan\left(\frac{\t+\e(\t)}{2}\right)\ .
\eal
Keeping terms upto quadratic order in the fields, we get
\bal
I^{\mathcal{N}=2}=&\frac{1}{2}\int d\t \left(\e '(\t)^2	-\e ''(\t)^2-4 a'(\t)^2-8 \h (\t)  \bar\h^{(3)}(\t)-2 \h (\t)  \bar\h'(\t)+1\right)
\eal
Comparing with the answer in~\eqref{schker}, we find exact agreement with the $\mathcal{N} = 2$ super-Schwarzian derived from our ladder kernel expressions, up to an overall coefficient that is not determined in this approach.

%%%%%%%%%%%%%%%%%%%%%%%%%%
\section{The ground state contributions to the Green's function}\label{gd}
%%%%%%%%%%%%%%%%%%%%%%%%%%
It is shown in~\cite{Fu:2016vas,Stanford:2017thb} that the $\mathcal{N}=2$ SUSY SYK model in $0+1d$ has exact ground states. In this section we want to understand better the properties of these zero modes and in particular their contributions to the Green's function. 

We consider the low energy spectral density of the model worked out in~\cite{Stanford:2017thb}
\bal
\r_n(E)= \frac{\cos(\pi n q )}{1-4n^2 q^2}\left[\delta(E)+\sqrt{\frac{a_n}{E}}I_1(2 \sqrt{{a_n}{E}})\right]\,,\qquad a_n=2\p^2 C(1-4n^2 q^2)\,,
\eal
where $n$ is the charge under the $u(1)$ and $I_1$ is the modified Bessel function of the first kind. This is simply the Laplacian transform of the partition function in the Schwarzian effective theory
\bal
\int dE e^{-\b E} \r_n(E)=\frac{\cos(\pi n q )}{1-4n^2 q^2} e^{\frac{a_n}{\b}}\ . \label{laprho}
\eal

\begin{figure}[t!]
	\centering
	\begin{subfigure}[t]{0.49\textwidth}
		\centering
		\includegraphics[width=0.96\linewidth]{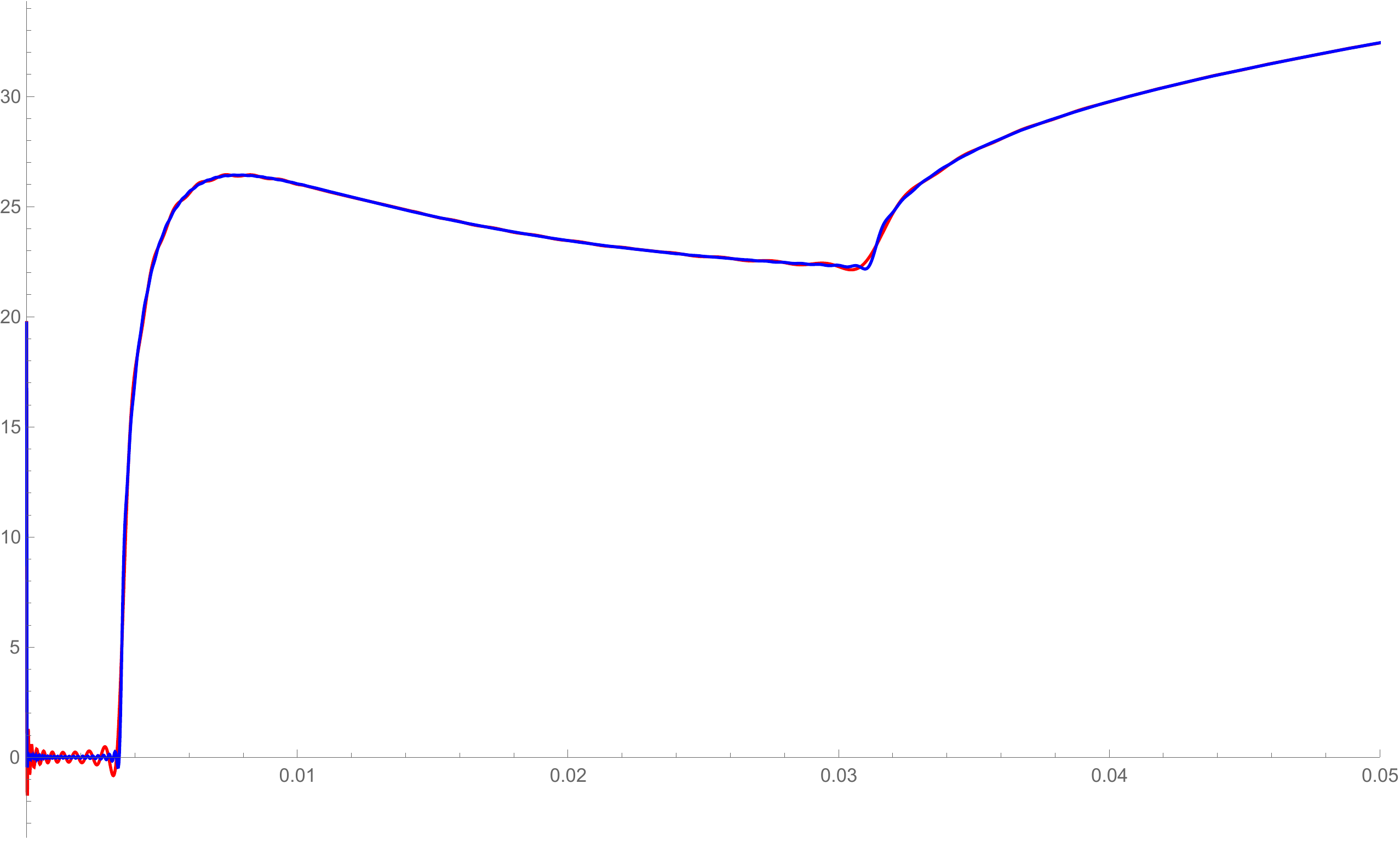}
		\caption{The spectral density for even $N$. The red and blue curve are $\L=20$ and $\L=60$ respectively. The plot is computed at $q=3$, $C=1$.}\label{fig:rhoeven}
	\end{subfigure}%
	~ 
	\begin{subfigure}[t]{0.49\textwidth}
		\centering
		\includegraphics[width=0.96\linewidth]{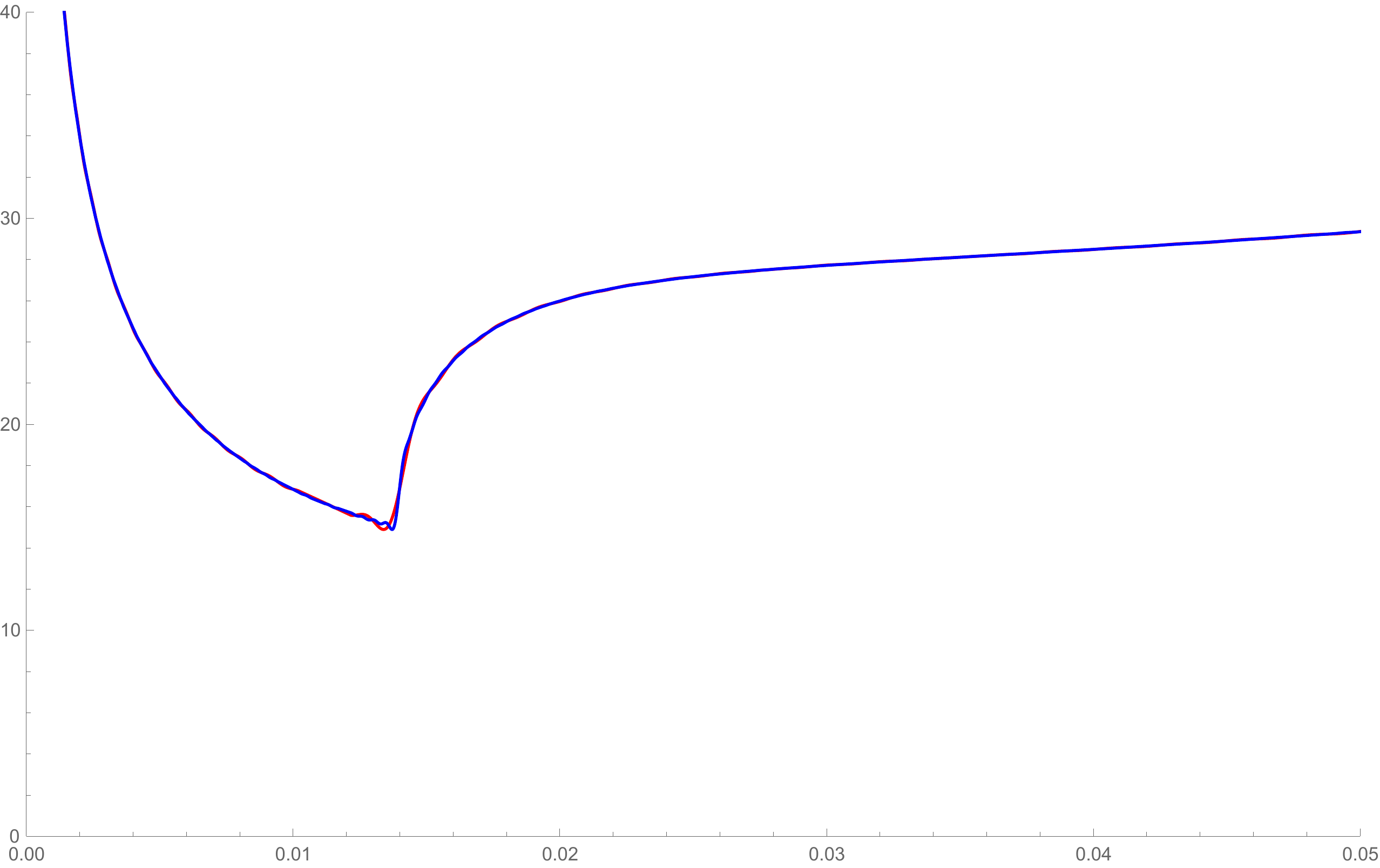}
		\caption{The spectral density for odd $N$. The red and blue curve are $\L=20$ and $\L=60$ respectively. The plot is computed at $q=3$, $C=1$.}\label{fig:rhoodd}
	\end{subfigure}
	\caption{The spectral density of the $\mathcal{N}=2$ model.}
\end{figure}

The full density matrices are 
\bal
\r_{\text{even}}(E)&=\sum_{n=-\infty}^\infty \r_n(E)\,,\qquad  
\r_{\text{odd}}(E)=\sum_{n=-\infty}^\infty (-1)^n \r_n(E)\,,
\eal  
where the subscript ``even" and ``odd" label the oddity of $N$.
Formally, we can carry out the sum to get
\bal
\r_{o}(E)=C^o_1 \delta(E) + C^o_2(E)\,,\qquad o=\text{ even or odd}\ .\label{rhogen1}
\eal
we get
\bal
C^{\text{even}}_1=\frac{\p}{2 q \sin(\frac{\p}{2q}) }\,,\qquad C^{\text{odd}}_1=\frac{\p}{2 q }\cot(\frac{\p}{2q})\ .
\eal
Notice that in the large-$q$ limit the coefficients are
\bal
\lim\limits_{q\to \infty }C_1^{\text{even}}&=1\,,\qquad 
\lim\limits_{q\to \infty }C_1^{\text{odd}}=1\ .
\eal
While the full $C^o_2(E)$ function is complicated to get explicitly, we can   carry out this summation numerically  and the results are shown in Fig~\ref{fig:rhoeven} and Fig~\ref{fig:rhoodd}. 

In particular we observe
\bal
C_{2}^{\text{even}}(0)&=2\p^2\,,\qquad  C_{2}^{\text{odd}}(0)= 2 \pi ^2 (2 \Lambda +1)\,,
\eal
where $\Lambda$ is the UV cutoff that should be taken to $\infty$.  

We want to understand the effect of the exact ground states, represented by the $\delta(E)$ function, in the above density matrix. 

We start with the Euclidean 2-point function
\bal
G^E_{AB}(\t)&=\langle A(\t)B(0)\rangle =\langle e^{H\t}A(0)e^{-H\t}B(0)\rangle \\
&=\text{Tr} \left[\sum_m |m\rangle \langle m| e^{-\b H }e^{H\t}A(0)e^{-H\t}B(0)\right]\\
&=\text{Tr} \left[\sum_{m,n} |m\rangle \langle m| \left(e^{(\t-\b) E_m-E_n\t }A(0)  |n\rangle \langle n|B(0)\right)\right]\\
&\sim\frac{1}{\mathcal{N} ^2}\int dE_m dE_n \r(E_m)\r(E_n)  \left(e^{(E_m-E_n)\t-E_m\b }A_{nm} B_{mn}\right)
\eal
where $A_{mn}$ and $B_{mn}$ are matrix elements and $\mathcal{N} $ is the normalization factor
\bal
\mathcal{N} =\int d E_m \r(E_m)\ . 
\eal 
Further notice that the sign $\sim$ is to emphasis that here we approximate the two sums by two separate integrals over the density of states. 

\subsection{Energy-energy correlator}

We can start to consider a special correlator that can be computed exactly. 
Consider the energy-energy correlator
\bal
\langle H(\t)H(0)\rangle &=\int dE_m dE_n \r(E_m)\r(E_n)  e^{(E_m-E_n)\t-E_m\b }E^2_{n} \delta_{m,n}\\
&=\int dE_m \r(E_m)^2 E_{m}^2 e^{-E_m\b }\\
&=\int dE_m C_2(E_m)^2 E_{m}^2 e^{-E_m\b }\ .
\eal
So it only revceives contribution from the continuous spectrum, this means some special correlation functions does not receive contributions from the exact ground states of the $\mathcal{N}=2$ model.

%%%%%%%%%%%%%%%%%%%%%%%
\subsection{A crude approximation}
%%%%%%%%%%%%%%%%%%%%%%%

We can consider more general correlation funcitons. To evaluate those, in this section we would like to use a crude approximation inspired from the ETH hypothesis~\cite{Deutsch1991,Srednicki1994} 
\bal
A_{mn}=a \delta_{m,n}\,,\qquad B_{mn}=b \delta_{m,n}\ .
\eal
Then the above propagator reduces to
\bal
G^E_{AB}&=ab \int dE_m  \r(E_m)^2 e^{-E_m\b }\\
&=ab \int dE_m  \left(C_1^2\delta(E_m)+2C_1C_2(E_m) \delta(E_m) +C_2(E_m)^2\right) e^{-E_m\b }\\
&=ab  \left[ C_1^2 +2C_1C_2(0) +\int dE_m  C_2(E_m)^2 e^{-E_m\b }\right]\,,
\eal
where we have used
\bal
\r(E)=C_1 \delta(E) + C_2(E)\,,
\eal
for $N$ being either even or odd.

We can compare the relative size of the contributions from continuous spectrum and the ground states. 
At very low temperature $\b \to \infty$, the integral is localized at $E_m=0$.  We observe that the contribution from the continuous spectrum is much larger than the contribution from the grounds state
\bal
\frac{C_2(0)^2}{C_1^2+2 C_1 C_2(0)} \to \infty\ .
\eal 
It is clear that as the temperature increases more and more high energy modes contribute significantely so the contribution from the continuous spectrum is more and more dominant.

It is true that we are using a very crude aproximation so the above statement might be too extremal, but we  will see in our later less crude approximation that this is a general property.

\subsection{A better approximation}

We can consider the case where the matrix elements are assumed to be constant, 
\bal
A_{mn}=a \,,\qquad B_{mn}=b \,,
\eal
which is the opposite extreme of assuming them to be diagonal.
Then the correlation function reads
\bal
G^E_{AB}(\t)&\sim\frac{1}{\mathcal{N} ^2}\int dE_m dE_n \r(E_m)\r(E_n)  e^{(E_m-E_n)\t-E_m\b }A_{nm} B_{mn}\\
&=  \frac{ab}{\mathcal{N} ^2}\int dE_m  e^{-(\b-\t) E_m}\r(E_m)\int dE_n   \r(E_n) e^{-\t E_n  }\\
&=  \frac{ab}{\mathcal{N} ^2} Z(\b-\t)Z(\t)\ .
\eal
To proceed, we can the use the explicit form of the partition functions
\bal
Z^{\text{even}}(\b)&=\int dE e^{-\b E} \r^{\text{even}}(E)=\sum_{n=-\infty}^{\infty}\int dE e^{-\b E} \r^{\text{even}}_n(E)\\
&=\sum_{n=-\infty}^{\infty}\frac{\cos(\pi n q )}{1-4n^2 q^2} e^{\frac{2\p^2 C(1-4n^2 q^2)}{\b}}\\
&=\int dC \frac{2\p^2}{\b} \sum_{n=-\infty}^{\infty}\cos(\pi n q )e^{\frac{2\p^2 C(1-4n^2 q^2)}{\b}}\\
&=\int dy \,\left(\vartheta _3\left(0,y^{-16 q^2}\right)-\vartheta _2\left(0,y^{-16 q^2}\right)\right)\,,\label{zeven}
\eal
and for the odd $N$ case
\bal
Z^{\text{odd}}(\b)&=\int dE e^{-\b E} \r^{\text{odd}}(E)=\sum_{n=-\infty}^{\infty}\int dE e^{-\b E} \r^{\text{odd}}_n(E)\\
&=\sum_{n=-\infty}^{\infty}\frac{1}{1-4n^2 q^2} e^{\frac{2\p^2 C(1-4n^2 q^2)}{\b}}\\
&=\int dC \frac{2\p^2}{\b} \sum_{n=-\infty}^{\infty}e^{\frac{2\p^2 C(1-4n^2 q^2)}{\b}}\\
&= \int dy \,\vartheta _3\left(0,y^{-4 q^2}\right)\ .\label{Zodd}
\eal
To separate the ground states contribution from the continuous spectrum contribution we compute seperately the contributions to the partition functions from the ground state 
\bal
I_g^{\text{odd}}&=\int dE e^{-\b E} C_1^{\text{odd}}\delta(E)=C_1^{\text{odd}}\\
I_g^{\text{even}}&=\int dE e^{-\b E} C_1^{\text{even}}\delta(E)=C_1^{\text{even}}\ .
\eal
The contributions from the continuous piece are
\bal
I_c^{\text{odd}}(\b)&=I^{\text{odd}}(\b)-C_1^{\text{odd}}\\
I_c^{\text{even}}(\b)&=I^{\text{even}}(\b)-C_1^{\text{even}}\ .
\eal

\begin{figure}
	\centering
	\includegraphics[width=0.95\linewidth]{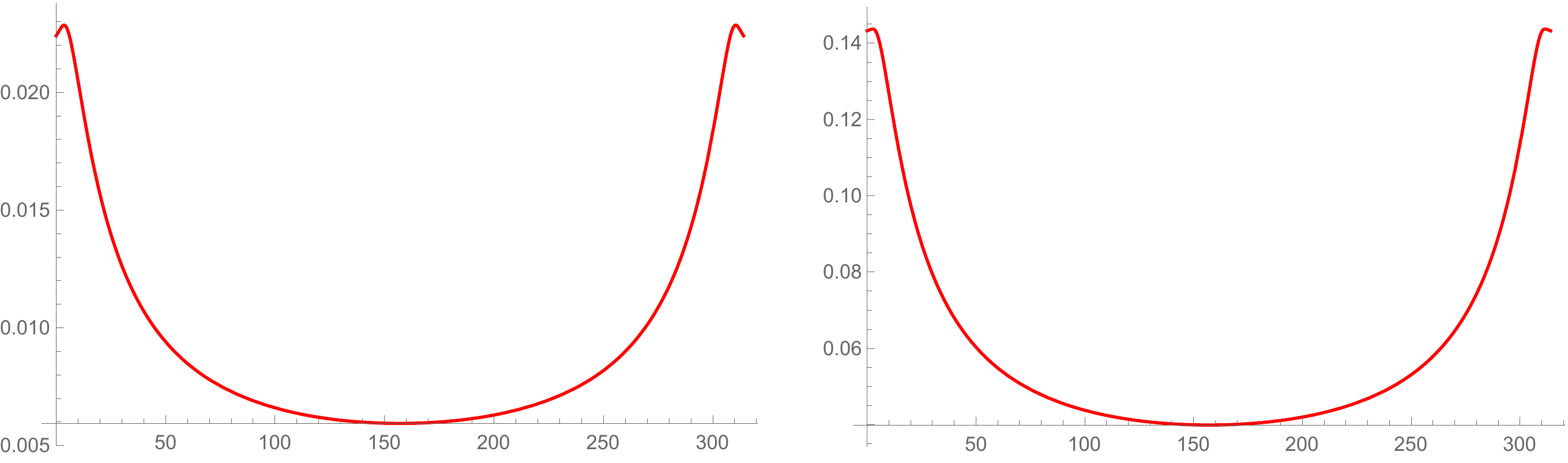}
	\caption{Contribution to the Green's function from the continuous spectrum. The computation is done at $\b = 100 \p$, $C=1$. The left panel is for the even $N$ case; the right panel is for the odd $N$ case. The horizontal axis is the Euclidean time ranging from 0 to $100\p$. The vertical axis is the contribution to the Green's function from the continuous spectrum in percentage. This is the temperature that is relevant to the discussion in~\cite{Berkooz:2020xne}}
	\label{fig:ratio400pi}
\end{figure}

\begin{figure}
	\centering
	\includegraphics[width=0.95\linewidth]{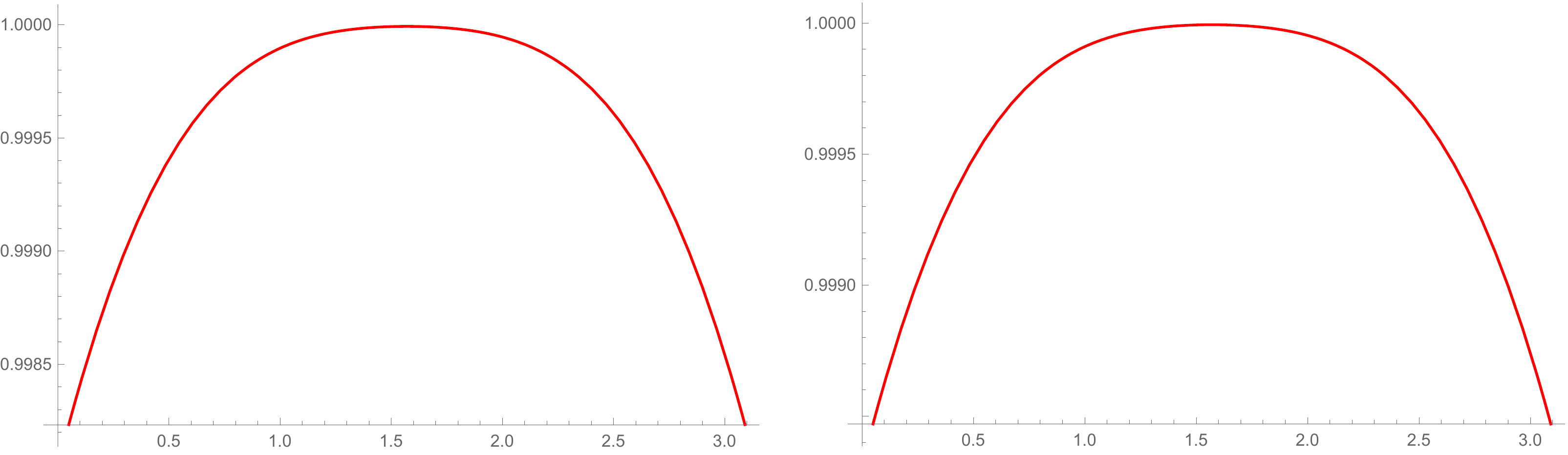}
	\caption{Percentage of the contribution to the Green's function from the continuous spectrum. The computation is done at $\b = \p$, $C=1$.  The left panel is for the even $N$ case; the right panel is for the odd $N$ case. }
	\label{fig:ratio2pi}
\end{figure}

We can now compare the contributions from the conformal continuous spectrum with the contributions from the ground states by numerically compute the two contributions to the Green's function. The answer is shown in Fig.~\ref{fig:ratio400pi} and Fig.~\ref{fig:ratio2pi}. There we see explicitly that as $\b$ decreases to relatively small value, the contribution from the continuous spectrum becomes more and more dominant.

Here we discuss a few relations between our results with that in~\cite{Berkooz:2020xne}. 
One conclusion of~\cite{Berkooz:2020xne} is that the contribution to the 2-point function from the ground state is never negligible in the douple scaling large-$q$ limit. From our previous computation we see this statement depends on the operators in the correlators; for example the ground state does not contribute to the energy-energy correlators. 

For a generic correlator, at very low temperature the high energy modes are not significantly excited so it is expected that the ground state contribution is always significant. This is the temperature range discussed in~\cite{Berkooz:2020xne}: the double scaling limit result, although correctly reproduce the energy spectrum, is valid in the range $1/\lambda \ll \beta \ll 1/\lambda^2$. So in the double scaling limit, namely $\lambda \to 0$, the inverse temperature $\beta$ is constrained to be infinite. As a result, the conclusion of~\cite{Berkooz:2020xne} that in the low energy domain the ground state contribution is always important is indeed consistent with our expectation.

On the other hand, as the temperature increases (but still in the range of the IR conformal window) one can check that the ground state contribution is less and less important. Eventually it becomes negligible and one can show that the dominant contribution is from the conformal spectrum. So in that wider temperature range the conformal answer remains a good approximation of the exact solution. This is what we explained above and have shown in Figure~\ref{fig:ratio400pi} and~\ref{fig:ratio2pi}. 

So at very low temperature the ground state contribution is always not negligible, while in a higher temperature range the conformal answer dominates and is a good approximation to the exact solution. Therefore although the conformal answer at zero temperature indeed only captures a finite piece of the exact 2-point function, but it is still special and useful in the sense that the finite temperature solutions can be obtained from it by a simple reparameterization and the result does contribute dominantly at finite temperatures.

%%%%%%%%%%%%%%%%%%%%%%%%%%%
\section{Schwarzian correlators}\label{SC}
%%%%%%%%%%%%%%%%%%%%%%%%%%%%

In this section we consider the correlators of the super-Schwarzian operator~\eqref{ssch2},~\eqref{ssch1} in the supersymmetric model such as
\bal
\langle S^b({\bm  \t}_1)S^b( {\bm\t}_2)\ldots S^b( {\bm\t}_n)\rangle\,,
\eal
where we have used the short hand notaton $S^b({\bm  \t})=S^b\left(\t_1,\q_1,\bar{\q}_1;\t,\q,\bar{\q}\right)$.

One crucial difference between our following computation in the supersymmetric model from the Schwarzian correlators in the pure fermionic model computation is that the $\mathcal{N}=2$ model has a $U(1)$ symmetry as well as fermionic (bilinear) components in the $\mathcal{N}=2$ superconformal algebra and the partition function receives contributions form different winding sectors. This can be seem from the explicit expression of the partition function~\eqref{zeven} and~\eqref{Zodd}. For later convenience, we first recast them together  with their derivatives here. As obtained in~\cite{Stanford:2017thb}, the partition function for any given winding number $n$ is again one-loop exact and can be written as
\bal
Z_{n}(\beta)=\frac{\cos (\pi {q} n)}{1-4 {q}^{2} n^{2}} \exp \left[\frac{2 \pi^{2} C}{\beta}\left(1-4 n^{2} {q}^{2}\right)\right]\ .
\eal 
The full partition function is then
\bal
Z^{\text{even}} =\sum_{n=-\infty}^\infty Z_n(\b)\,,
\qquad
Z^{\text{odd}} =\sum_{n=-\infty}^\infty (-1)^n Z_n(\b)\ .
\eal
Analytic (integral) expressions of $Z^{\text{ even/ odd}}$ can be found in~\eqref{zeven} and~\eqref{Zodd}, we can further get its derivative
\bal
\frac{\pa}{\pa \tilde{g}} Z^{\text{even}} &=\frac{\pa}{\pa \tilde{g}}\sum_{n=-\infty}^{\infty}\frac{\cos(\pi n q )}{1-4n^2 q^2} e^{\tilde{g} \p (1-4n^2 q^2)}=\p\sum_{n=-\infty}^{\infty}{\cos(\pi n q )} e^{\tilde{g} \p (1-4n^2 q^2)}\\
&=\pi e^{\pi  \tilde{g}} \left(\vartheta _3\left(0,e^{-16 \tilde{g} \pi  q^2}\right)-\vartheta _2\left(0,e^{-16 \tilde{g} \pi  q^2}\right)\right)\,,\label{pzeven}
\eal 
and
\bal
\frac{\pa}{\pa \tilde{g}} Z^{\text{odd}} &=\frac{\pa}{\pa \tilde{g}}\sum_{n=-\infty}^{\infty}\frac{1}{1-4n^2 q^2} e^{\tilde{g} \p (1-4n^2 q^2)}=\p\sum_{n=-\infty}^{\infty} e^{\tilde{g} \p (1-4n^2 q^2)}\\
&=\pi e^{\pi  \tilde{g}}\vartheta _3\left(0,e^{-4 \tilde{g} \pi  q^2}\right)\,,\label{pzodd}
\eal 
where
\bal
\tilde{g}=\frac{2\p C}{\b}\ .
\eal
Further derivatives can be act on the theta functions. 

Another crucial difference is the property of the super-Schwarzian operator
\bal
S\left(\tau_1, \theta_1, \bar{\theta}_1 ; \tau, \theta, \theta\right)=\frac{\partial_\t D \bar{\theta}}{D \theta_1}-\frac{\partial_\t D \theta_1}{D \theta_1}-2 \frac{\partial_{\t} \theta_1 \partial_{\t} \bar{\theta}_1}{\left(D \bar{\theta}_1\right)\left(D \theta_1\right)} \,, 
\eal
which we now review. 
The super-Schwarzian satisfies
\bal
S\left(\tau_2, \theta_2, \bar{\theta}_2 ; \tau, \theta, \bar{\theta}\right)=\left(D \theta_1\right)\left(\bar{D} \bar{\theta}_1\right) S\left(\tau_2, \theta_2, \bar{\theta}_2 ; \tau_1, \theta_1, \bar{\theta}_1\right)+S\left(\tau_1, \theta_1, \bar{\theta}_1 ; \tau, \theta, \bar{\theta}\right)\ .\label{schrec}
\eal
For simplicity, we rewrite the super-Schwarzian derivative operation in terms of the maps ${\bf f}_1, {\bf f}_2 \in \mathbb{R}^{1|2} \rightarrow  \mathbb{R}^{1|2}$:
\bal
{\bf f}_1: &\quad (\t,\q,\bar{\q}) \mapsto \left(\t_1,\q_1,\bar{\q}_1\right)\\
{\bf f}_2: &\quad (\t_1,\q_1,\bar{\q}_1) \mapsto \left(\t_2,\q_2,\bar{\q}_2\right)\ .
\eal
For example 
$S\left(\tau_1, \theta_1, \bar{\theta}_1 ; \tau, \theta, \bar{\theta}\right)$ associated to the above transformation can be denoted by $S\left({\bf f}_1\right)$. The above relation can be recast into the form
\bal
S({\bf f}_2\circ {\bf f}_1)=\left(D \theta_1\right)\left(\bar{D} \bar{\theta}_1\right) S({\bf f}_2)\circ {\bf f}_1+S({\bf f}_1)\ .
\eal  
As a consequency, we get
\bal
&0=S({\bf f}_1^{-1}\circ {\bf f}_1)=\left(D \theta_1\right)\left(\bar{D} \bar{\theta}_1\right) S({\bf f}_1^{-1})\circ {\bf f}_1+S({\bf f}_1)\\
&\Rightarrow \qquad S({\bf f}_1)=-\left(D \theta_1\right)\left(\bar{D} \bar{\theta}_1\right) S({\bf f}_1^{-1})\circ {\bf f}_1\ .
\eal
In the coordinate basis, this gives the inversion formula
\bal
\qquad S(\t_1,\q_1,\bar{\q}_1;\t,\q,\bar{\q})=-\left(D \theta_1\right)\left(\bar{D} \bar{\theta}_1\right) S(\t,\q,\bar{\q};\t_1,\q_1,\bar{\q}_1)\ .\label{inv}
\eal

The full partition function of the supersymmetric Schwarzian theory is~\cite{Stanford:2017thb}
\bal
&Z=\int \frac{\mathcal{D} \f_1 \mathcal{D}\s \mathcal{D}\h_1 \mathcal{D}\bar\h}{Osp(2|2)} \exp \left(I^{\mathcal{N}=2} \right)\equiv \int d\mu[{\bf f}_1] \exp\left(I^{\mathcal{N}=2}\right)\,,
\eal
where 
\bal
I^{\mathcal{N}=2}&=\tilde{g}\int_{0}^{2 \pi} d \tau \int d\q d\bar\q S\left(\tau_1, \theta_1, \bar{\theta}_1 ; \tau, \theta, \bar{\theta}\right) =\tilde{g}\int_{0}^{2 \pi} d \tau  S^{b}\left(\tau_1, \theta_1, \bar{\theta}_1 ; \tau, \theta, \bar{\theta}\right)\,,
\eal
with $S^{b}\left(\tau_1, \theta_1, \bar{\theta}_1 ; \tau, \theta, \bar{\theta}\right)$ purely bosonic as defined in~\eqref{ssch2},~\eqref{ssch1} and the $(\t_1,\q_1,\bar{\q}_1)$ is a super-reparameterization of the super-coordinates. The super-reparameterization is requred to preserve the chirality of the two supersymmetries, this means the super-derivatives obey 
\bal
{D}_\q \bar\q' = 0\,,\qquad {D}_{\bar\q} \q' = 0\,,\qquad  {D}_\q \t' = \bar{\q}' D_\q \q'\,,\qquad \bar{D}_{\bar\q} \t' = {\q}' \bar D_{\bar\q} \bar\q'\,,\label{chiralcst}
\eal
where we have spelt out the $\q $ and $\bar\q$ in the super-derivative for clarification. 

Although we will not use the explicit parameterization of the super-reparameterization transformation, for illustration purpose we provide one example~\cite{Fu:2016vas}
\bal
\t' &= f(\t)+\q \bar{g}(\t)+\bar{\q} {g}(\t)+\q \bar\q h(\t)\\
\q'&=\r(\t+\q\bar{\q})\left(\q+\h(\t+\q\bar{\q})\right)\\
\bar\q'&=\bar\r(\t-\q\bar{\q})\left(\q+\bar\h(\t-\q\bar{\q})\right)\,,
\eal 
where
\bal
\r(\t)&= e^{i \sigma (t)} {f'(t)}^{\frac{1}{2}} \left(1+\frac{1}{2} \eta (t)  \bar{\eta}'(t)+\frac{1}{2}  \bar{\eta}(t) \eta '(t)+ i  \bar{\eta}(t) \eta (t) \sigma '(t)+\frac{3}{4}  \bar{\eta}(t) \eta (t)  \bar{\eta}'(t) \eta '(t)\right)\\
\bar{\r}(\t)&= e^{-i \sigma (t)} {f'(t)}^{\frac{1}{2}} \left(1+\frac{1}{2} \eta (t)  \bar{\eta}'(t)+\frac{1}{2}  \bar{\eta}(t) \eta '(t)+ i  \bar{\eta}(t) \eta (t) \sigma '(t)+\frac{3}{4}  \bar{\eta}(t) \eta (t)  \bar{\eta}'(t) \eta '(t)\right)\\
g(\t)&=\r(\t)\bar{\r}(\t)\h(\t)\\
\bar{g}(\t)&=\r(\t)\bar{\r}(\t)\bar{\h}(\t)\\
h(\t)&=\r(\t)\h(\t)\pa_\t \left(\bar{\r}(\t)\bar{\h}(\t)\right)-\bar\r(\t)\bar\h(\t)\pa_\t \left({\r}(\t){\h}(\t)\right)\ .
\eal

With all the prepration, we now proceed to compute the correlators of the supery-Schwarzian operator, following a general method discussed in~\cite{Stanford:2017thb}. Using the relation~\eqref{schrec}, we can compute
\bal
Z^{\mathcal{N}=2}&=\int d\mu[{\bf f}_2] \exp\left(\tilde{g}\int d\t d\q d\bar\q S({\bf f}_2)\right)\\
&=\int d\mu[{\bf f}_2] \exp\left(\tilde{g}\int d\t d\q d\bar\q \left(D \theta_1 \bar{D} \bar{\theta}_1 S\left(\tau_2, \theta_2, \bar{\theta}_2 ; \tau_1, \theta_1, \bar{\theta}_1\right)+S\left({\bf f}_1\right)\right)\right)\ .
\eal
Because the second term in the exponential does not depend on ${\bf f}_2$, its integration over the measure is trivially identity, we can thus rewrite the above into
\bal
&\int d\mu[{\bf f}_2] \exp\left(\tilde{g}\int d\t d\q d\bar\q \left(D \theta_1 \bar{D} \bar{\theta}_1 S\left(\tau_2, \theta_2, \bar{\theta}_2 ; \tau_1, \theta_1, \bar{\theta}_1\right)\right)\right)\\
&=e^{-\tilde{g}\int d\t d\q d\bar\q S\left({\bf f}_1\right)}Z^{\mathcal{N}=2}=\exp{\left(\tilde{g}\int d\t d\q d\bar\q \,\left(D \theta_1 \bar{D} \bar{\theta}_1 S\left(\tau, \theta, \bar{\theta}; \tau_1, \theta_1, \bar{\theta}_1\right)\right)\right)}Z^{\mathcal{N}=2}\ .\label{rel1}
\eal
Next we change the integral measure according to
\bal
d\t_1 d\q_1 d \bar{\q}_1=\text{Ber}\left(\frac{\pa(\t_1,\q_1,\bar{\q}_1)}{\pa(\t,\q,\bar{\q})}\right) d\t d\q d \bar{\q} 
\eal 
where $\text{Ber}$ stands for the Berezinian of the transformation Jacobian
\bal
J=\left(\frac{\pa(\t_1,\q_1,\bar{\q}_1)}{\pa(\t,\q,\bar{\q})}\right)&=\begin{pmatrix}
	\pa_{\t}\t_1 & \pa_{\t}\q_1 & \pa_{\t}\bar\q_1 \\
	\pa_{\q}\t_1 & \pa_{\q}\q_1 & \pa_{\q}\bar\q_1\\
	\pa_{\bar\q}\t_1 & \pa_{\bar\q}\q_1 & \pa_{\bar\q}\bar\q_1
\end{pmatrix}\ .
\eal
Making use of the definition
\bal
D=\partial_\q +\bar{\q} \partial_{\t}\,,\qquad \bar{D}=\partial_{\bar\q}+\q \partial_{\t}\,,
\eal
we rewrite the above Jacobian factor into
\bal
J&=\begin{pmatrix}
	\pa_{\t}\t_1 & \pa_{\t}\q_1 & \pa_{\t}\bar\q_1 \\
	\pa_{\q}\t_1 & \pa_{\q}\q_1 & \pa_{\q}\bar\q_1\\
	\pa_{\bar\q}\t_1 & \pa_{\bar\q}\q_1 & \pa_{\bar\q}\bar\q_1
\end{pmatrix}=\begin{pmatrix}
	\pa_{\t}\t_1 & \pa_{\t}\q_1 & \pa_{\t}\bar\q_1 \\
	D\t_1 & D\q_1 & D\bar\q_1\\
	\bar{D}\t_1 & \bar{D}\q_1 & \bar{D}\bar\q_1
\end{pmatrix}=\begin{pmatrix}
\pa_{\t}\t_1 & \pa_{\t}\q_1 & \pa_{\t}\bar\q_1 \\
\bar{\q}_1 D\q_1 & D\q_1 & 0\\
\q_1 \bar{D}\bar\q_1 & 0 & \bar{D}\bar\q_1
\end{pmatrix}\ .
\eal

For a supermatrix with the block form
\bal
X=\begin{pmatrix}
	A & B\\
	C & D
\end{pmatrix}\,,
\eal
the Berezinian exists if both $A$ and $D$ are invertible, in which case the Berezinian is defined by
\bal
\text{ Ber }  (X)=\det(A-BD^{-1} C)\det(D^{-1})\ .
\eal 
With this we can evaluate the Berenzinian as
\bal
\text{ Ber }  (J)&= \text{ Ber }\begin{pmatrix}
	\pa_{\t}\t_1 & \pa_{\t}\q_1 & \pa_{\t}\bar\q_1 \\
	\bar{\q}_1 D\q_1 & D\q_1 & 0\\
	\q_1 \bar{D}\bar\q_1 & 0 & \bar{D}\bar\q_1
\end{pmatrix}=(D\q_1)^{-1}(\bar{D}\bar\q_1)^{-1}\text{ Ber }\begin{pmatrix}
\pa_{\t}\t_1 & \pa_{\t}\q_1 & \pa_{\t}\bar\q_1 \\
\bar{\q}_1  & 1 & 0\\
\q_1  & 0 & 1
\end{pmatrix}\\
&=(D\q_1)^{-1}(\bar{D}\bar\q_1)^{-1}\left(\pa_{\t}\t_1+\bar\q_1\pa_{\t}\q_1 +{\q}_1 \pa_{\t}\bar\q_1\right)=1\,,
\eal
where we have used~\eqref{chiralcst}
\bal
\{D,\bar{D}\}=2\pa_\t\,,
\eal
to rewrite
\bal
\pa_{\t}\t_1+\bar\q_1\pa_{\t}\q_1 +{\q}_1 \pa_{\t}\bar\q_1=(D\q_1) (\bar{D}\bar{\q}_1)\ .
\eal
We can thus rewrite~\eqref{rel1} as
\bal
&\int d\mu[{\bf f}_2] \exp\left(\tilde{g}\int d\t_1 d\q_1 d\bar\q_1 \left(D \theta_1 \bar{D} \bar{\theta}_1 S\left(\tau_2, \theta_2, \bar{\theta}_2 ; \tau_1, \theta_1, \bar{\theta}_1\right)\right)\right)\\
&=e^{-\tilde{g}\int d\t d\q d\bar\q S\left({\bf f}_1\right)}Z^{\mathcal{N}=2}=\exp{\left(\tilde{g}\int d\t_1 d\q_1 d\bar\q_1 \,\left(D \theta_1 \bar{D} \bar{\theta}_1 S\left(\tau, \theta, \bar{\theta}; \tau_1, \theta_1, \bar{\theta}_1\right)\right)\right)}Z^{\mathcal{N}=2}\ .
\eal
Now we can choose a special function ${\bf f}_1: \{\t,\q,\bar\q\}\mapsto\{\t_1,\q_1,\bar{\q}_1\}$ 
\bal
\t_1=\t+\e(\t)\,, \qquad \q_1=(1+\pa\e(\t))^{\frac{1}{2}}\q\,,\qquad \bar{\q}_1=(1+\pa\e(\t))^{\frac{1}{2}}\bar\q\ .
\eal
It is easy to find its invese $\left({\bf f}_1\right)^{-1}$ 
\bal
&\t=\t_1-\e_1(\t_1)\,, \qquad \e_1(\t)=\e(\t-\e(\t))\\ &\q=\left(1+\frac{\pa_{\t_1}\e(\t_1)}{1-\pa_{\t_1}\e_1(\t_1)}\right)^{-\frac{1}{2}}\q_1\,,\quad \bar{\q}=\left(1+\frac{\pa_{\t_1}\e(\t_1)}{1-\pa_{\t_1}\e_1(\t_1)}\right)^{-\frac{1}{2}}\bar\q_1\ .
\eal
Then from above we get
\bal
&\int d\mu[{\bf f}_2] \exp\left(\tilde{g}\int d\t_1 d\q_1 d\bar\q_1 \left(D \theta_1 \bar{D} \bar{\theta}_1 S\left(\tau_2, \theta_2, \bar{\theta}_2 ; \tau_1, \theta_1, \bar{\theta}_1\right)\right)\right) \label{firstline}\\
&=e^{-\tilde{g}\int d\t d\q d\bar\q S\left({\bf f}_1\right)}Z^{\mathcal{N}=2}=e^{\tilde{g}\int d\t \frac{\left(2 \epsilon ^{(3)}(t) \left(\epsilon '(t)+1\right)-3 \epsilon ''(t)^2\right)}{2 \left(\epsilon '(t)+1\right)^2}}Z^{\mathcal{N}=2}\ .
\eal
We can further evaluate  the factor $D \theta_1 \bar{D} \bar{\theta}_1 $
\bal
D \theta_1 \bar{D} \bar{\theta}_1&=\left(\frac{-\q \bar{\q}  \epsilon ''(\t)}{2 \sqrt{\epsilon '(\t)+1}}+\sqrt{\epsilon '(\t)+1}\right)
\left(\frac{\q \bar{\q}  \epsilon ''(\t)}{2 \sqrt{\epsilon '(\t)+1}}+\sqrt{\epsilon '(\t)+1}\right)\\
&=1+\epsilon '(\t)=1+\frac{\epsilon (\t_1)}{\t_1}\frac{\t_1}{\t}=1+\frac{\pa_{\t_1}\e(\t_1)}{1-\pa_{\t_1}\e(\t_1-\e(\t_1))}\,,
\eal
with which we can expand the two sides of the equation in terms of the small quantity $\e(\t)$ and its derivatives, namely 
\bal
&\int d\mu[{\bf f}_2] \exp\left(\tilde{g}\int d\t_1 d\q_1 d\bar\q_1 \left(1+\frac{\pa_{\t_1}\e(\t_1)}{1-\pa_{\t_1}\e(\t_1-\e(\t_1))}\right) S\left(\tau_2, \theta_2, \bar{\theta}_2 ; \tau_1, \theta_1, \bar{\theta}_1\right) \right) \\
&=e^{\tilde{g}\int d\t \left(-\frac{3}{2}\left(\pa^2\e(\t)\right)^2\right)} Z^{\mathcal{N}=2}\ .\label{schrec2}
\eal
Expanding~\eqref{schrec2} to the leading order, the above relation leads to 
\bal
0&=\int d\mu[{\bf f}_2]\int d\t_1 d\q_1 d\bar\q_1 \pa_{\t_1}\e(\t_1) S\left(\tau_2, \theta_2, \bar{\theta}_2 ; \tau_1, \theta_1, \bar{\theta}_1\right)  e^{\tilde{g}\int d\t_1 d\q_1 d\bar\q_1  S\left(\tau_2, \theta_2, \bar{\theta}_2 ; \tau_1, \theta_1, \bar{\theta}_1\right) }\ .
\eal
Requiring periodicity in the time circle and using integration by parts, this leads to
\bal
&\langle \int d\t_1 d\q_1 d\bar\q_1 \e(\t_1)\pa_{\t_1} S\left(\tau_2, \theta_2, \bar{\theta}_2 ; \tau_1, \theta_1, \bar{\theta}_1\right)\rangle =0 \\
&\Rightarrow \langle S^b\left(\tau_2, \theta_2, \bar{\theta}_2 ; \tau_1, \theta_1, \bar{\theta}_1\right)\rangle =\text{ const }\,,
\eal
where
\bal
S^b\left(\tau_2, \theta_2, \bar{\theta}_2 ; \tau_1, \theta_1, \bar{\theta}_1\right)=\int d\q d\bar{\q} S\left(\tau_2, \theta_2, \bar{\theta}_2 ; \tau_1, \theta_1, \bar{\theta}_1\right)\,,
\eal
is the $\mathcal{N}=2$ physicsl Schwarzian operator and the constant can be obtained from
\bal
\int d\t_1 \langle S^b\left(\tau_2, \theta_2, \bar{\theta}_2 ; \tau_1, \theta_1, \bar{\theta}_1\right)\rangle=\frac{1}{Z^{\text{ even/ odd}}}\pa_{\tilde{g}}Z^{\text{ even/ odd}}\,,
\eal
whose right-hand-side can be simply evaluated by plugging in the results of~\eqref{pzeven} or~\eqref{pzodd} and~\eqref{zeven} or~\eqref{Zodd}. We spare the readers for the detailed expressions here since it's not very illuminating. 

Expanding~\eqref{schrec2} to the quadratic order, We have
\bal
\tilde{g}\langle S^b({\bm  \t})S^b( \tilde{\bm\t})\rangle +2\langle S^b({\bm\t})\rangle \delta(\t-\tilde{\t})&= -{3}  \pa_\t^2\delta(\t-\tilde{\t})  +\text{ const }\,,
\eal
where the constant term comes from the various integration by parts, and for simplicity we have used the short hand notaton $S^b({\bm  \t})=S^b\left(\t_1,\q_1,\bar{\q}_1;\t,\q,\bar{\q}\right)$. Like the one point function case, we can again determine the constants from
\bal
\int d\t d\tilde{\t} \langle S^b({\bm  \t})S^b(  \tilde{\bm\t})\rangle=\frac{1}{Z^{\text{ even/ odd}}}\pa^2_{\tilde{g}}Z^{\text{ even/ odd}}\,,
\eal
by plugging in the results of~\eqref{pzeven} or~\eqref{pzodd} and~\eqref{zeven} or~\eqref{Zodd}. We spare the readers for the detailed expressions here too. 

Higher point correlation functions can be obtained iteratively.

\section{Conclusions}

In this paper we consider varous properties of the contributions to the correlations functions of the low energy soft modes, including the degenerate exact ground states, in the $\mathcal{N}=2$ supersymmetric SYK model. We analyze the structure of the divergence of the 4-point correlation functions due to the stress tensor multiplet in the infrared, and regularize it by slightly going away from the conformal limit.  We also computed the chaotic exponent of this model away from the conformal fixed point in the large-$q$ limit and show that this channel gives the leading OTOC chaotic behavior~\cite{Shenker2014,Maldacena:2015waa}.  In addition, we derive the effective action of the stress tensor multiplet that correctly reproduces the large-contribution from the stress tensor multiplet to the 4-point function, and show that the resulting effective action precisely matches the previous result from the supersymmetrization of the Schwarzian action. We further derive the correlators of the Schwarzian operators.  We  comments on the contribution to the correlation functions from the ground states and show that it can be negligible for a large range of temperature, so that the zero temperature conformal solution is still useful in computations, for example to generate the conformal solutions at finite temperature. We also show explicitly that a second multiplet whose top component is another spin-2 operator, which could be present in the spectrum from previous analysis, does not actually appear in the spectrum in the conformal limit, and hence quantitatively answers the question about the existence of such a multiplet.  
Many of the discussions could also be applied to other SYK type systems, for example the coupled SYK model and its generalizations~\cite{Maldacena:2018lmt,Garcia-Garcia:2019poj, Alet:2020ehp,Plugge:2020wgc,Chen:2019qqe,Qi:2020ian, Sahoo:2020unu,py}. 

\section*{Acknowledgement}

We thank Luiz F. Alday, Micha Berkooz, Jordan Cotler, Vladimir Narovlansky, Mukund Rangamani, Zhenbin Yang and Zhixian Zhu for helpful discussion, and especially we are grateful to Yingfei Gu for enumarous interesting discussions and comments on an earlier version of this paper. CP would like to thank Shanghai Jiaotong University, Jilin University,  the workshop ``Higher Spin Gravity" workshop at the Asia Pacific Center for Theoretical Physics,  the Kavli Institute  for  Theoretical  Science  at  the  University  of  Chinese  Academy  of  Science,  Tsinghua University, the ``Interdisciplinary   workshop on Theoretical physics" at the Songshan Lake Materials Laboratory, University of Michigan, Ann Arbor, and the workshop ``Gauge Theories and Black Holes" at the Weizmann Institute of  Science  for  warm  hospitality  during  the  various  stages  of  the  project. 
The work of CP and SS was supported in part by the US Department of Energy under contract de-sc0010010 Task A. CP was also supported by the U.S. Department of Energy grant de-sc0019480 under the HEP-QIS QuantISED program and by funds from the Universityof California.

%%%%%%%
\appendix

\section{Details of the integrals to get the effective action}\label{integrals}

In the derivation of the Schwarzian effective action, we need to compute the inner  products of some bilocal functions. In this appendix we provide some useful formulas for this computation. 

First, the integral measure can be written as
\bal
\int_0^\beta d\t_1 d\t_2 =\int_{-\b}^{0}dx \int_{-\frac{x}{2}}^{\b+\frac{x}{2}} dy + \int^{\b}_{0}dx \int_{\frac{x}{2}}^{\b-\frac{x}{2}} dy \,,
\eal
where $y=\frac{\t_1+\t_2}{2}$ and $x=\t_1-\t_2$. For a special type of function
\bal
g(x,y)=e^{i \frac{2\p}{\b}(a-b) y} f_{a,b}(x)\,, \qquad a-b\in \mathbb{Z}\,,
\eal
the above integral simplifies to
\bal
\int_0^\beta d\t_1 d\t_2 e^{i \frac{2\p}{\b}(a-b) y} f_{a,b}(x) &=\int_{-\b}^{0}dx \int_{-\frac{x}{2}}^{\b+\frac{x}{2}} dy e^{i \frac{2\p}{\b}(a-b) y} f_{a,b}(x)\\
&\qquad\quad+ \int^{\b}_{0}dx \int_{\frac{x}{2}}^{\b-\frac{x}{2}} dy e^{i \frac{2\p}{\b}(a-b) y} f_{a,b}(x)\ .
\eal
If $a-b\neq 0$, we get
\bal
&\int_0^\beta d\t_1 d\t_2 e^{i \frac{2\p}{\b}(a-b) y} f_{a,b}(x) \\
&=\frac{1}{i \frac{2\p}{\b}(a-b)}\int_{-\b}^{0}dx  \left(e^{i \frac{2\p}{\b}(a-b) (\b+\frac{x}{2})}-e^{-i \frac{2\p}{\b}(a-b) \frac{x}{2}}\right) f_{a,b}(x)\\
&\qquad + \frac{1}{i \frac{2\p}{\b}(a-b)}\int^{\b}_{0}dx  \left(e^{i \frac{2\p}{\b}(a-b) (\b-\frac{x}{2})}-e^{i \frac{2\p}{\b}(a-b) \frac{x}{2}}\right) f_{a,b}(x)\\
&=\frac{1}{i \frac{2\p}{\b}(a-b)}\int^{\b}_{0}dx  \left(e^{i \frac{2\p}{\b}(a-b) (\b-\frac{x}{2})}-e^{i \frac{2\p}{\b}(a-b) \frac{x}{2}}\right) f_{a,b}(-x)\\
&\qquad + \frac{1}{i \frac{2\p}{\b}(a-b)}\int^{\b}_{0}dx  \left(e^{i \frac{2\p}{\b}(a-b) (\b-\frac{x}{2})}-e^{i \frac{2\p}{\b}(a-b) \frac{x}{2}}\right) f_{a,b}(x)\\
&=\frac{1}{i \frac{2\p}{\b}(a-b)}\int^{\b}_{0}dx  \left(e^{i \frac{2\p}{\b}(a-b) (\b-\frac{x}{2})}-e^{i \frac{2\p}{\b}(a-b) \frac{x}{2}}\right) \left( f_{a,b}(x)+f_{a,b}(-x)\right)\ .
\eal
Therefore as long as $ f_{a,b}(x)+f_{a,b}(-x)\neq 0$ the integral with $a-b\neq 0$ does not vanish. 

For the other case $a-b=0$, we get
\bal
\int_0^\beta d\t_1 d\t_2  f_{a,a}(x)& =\int_{-\b}^{0}dx \int_{-\frac{x}{2}}^{\b+\frac{x}{2}} dy f_{a,a}(x)+ \int^{\b}_{0}dx \int_{\frac{x}{2}}^{\b-\frac{x}{2}} dy  f_{a,a}(x)\\
&=\int_{-\b}^{0}dx {(\b+{x})}f_{a,a}(x)+ \int^{\b}_{0}dx (\b-{x}) f_{a,a}(x)\\
&= \int^{\b}_{0}dx (\b-{x}) (f_{a,a}(x)+f_{a,a}(-x))\ .
\eal

If we further sum over the discrete set of $a$, $b$ variable, for each pair of $(a, b)$ with $a-b\neq 0$, we get
\bal
&\sum_{a\neq b}\int_0^\beta d\t_1 d\t_2 e^{i \frac{2\p}{\b}(a-b) y} f_{a,b}(x) \\
&=\sum_{a}\sum_{b<a}\int_0^\beta d\t_1 d\t_2 \left(e^{i \frac{2\p}{\b}(a-b) y}+e^{i \frac{2\p}{\b}(b-a) y}\right) f_{a,b}(x) 
\eal
where we have assumed $f_{a,b}(x)=f_{b,a}(x)$. Using the above results, we get
\bal
&\sum_{a\neq b}\int_0^\beta d\t_1 d\t_2 e^{i \frac{2\p}{\b}(a-b) y} f_{a,b}(x) =\sum_{a}\sum_{ b<a}\frac{1}{i \frac{2\p}{\b}(a-b)}\\
&\times \int^{\b}_{0}dx  \left(e^{i \frac{2\p}{\b}(a-b) (\b-\frac{x}{2})}-e^{i \frac{2\p}{\b}(a-b) \frac{x}{2}}-e^{i \frac{2\p}{\b}(b-a) (\b-\frac{x}{2})}+e^{i \frac{2\p}{\b}(b-a) \frac{x}{2}}\right) \left( f_{a,b}(x)+f_{a,b}(-x)\right)
\eal
Because $a-b \in \mathbb{Z}$, we further get 
\bal
&\sum_{a\neq b}\int_0^\beta d\t_1 d\t_2 e^{i \frac{2\p}{\b}(a-b) y} f_{a,b}(x) =\sum_{a}\sum_{b<a}\frac{1}{i \frac{2\p}{\b}(a-b)}\\
&\times \int^{\b}_{0}dx  \left(e^{i \frac{2\p}{\b}(b-a) \frac{x}{2}}-e^{i \frac{2\p}{\b}(a-b) \frac{x}{2}}-e^{i \frac{2\p}{\b}(a-b) \frac{x}{2}}+e^{i \frac{2\p}{\b}(b-a) \frac{x}{2}}\right) \left( f_{a,b}(x)+f_{a,b}(-x)\right)\\
&=0\ .
\eal
Therefore the full summation localizes to the diagonal terms
\bal
&\sum_{a, b}\int_0^\beta d\t_1 d\t_2 e^{i \frac{2\p}{\b}(a-b) y} f_{a,b}(x)=\sum_{a}\int_0^\beta d\t_1 d\t_2  f_{a,a}(x)\\
&=\sum_{a}\int^{\b}_{0}dx (\b-{x}) (f_{a,a}(x)+f_{a,a}(-x))\ .
\eal
The concrete computations in the main text all follow the steps here, except for possible sign flips if necessary. The upshot of this computation is that we only need to compute the contribution from the diagonal entries in the summation.

\end{document}